\documentclass[english,a4paper,autoref,thm-restate,nameinlink,cleveref]{lipics-v2021}                          
\PassOptionsToPackage{svgnames}{xcolor}
\nolinenumbers
\hideLIPIcs

\ifpdf
  \pdfoutput=1\relax                    
  \usepackage{graphicx}                 
  \DeclareGraphicsExtensions{.pdf,.png,.jpg,.jpeg} 
\else
  \usepackage{graphicx}                 
  \DeclareGraphicsExtensions{.eps}      
\fi%

\graphicspath{{figures/}{pictures/}{images/}{./}} 
\usepackage{microtype}    
\PassOptionsToPackage{warn}{textcomp}   

\usepackage{cite}                       
\usepackage{tabu}                       
\usepackage{booktabs}                   

\usepackage{todonotes}
\usepackage[T1]{fontenc}
\PassOptionsToPackage{lineno}{switch}

\usepackage{colortbl}
\newcommand*{\belowrulesepcolor}[1]{%
  \noalign{%
    \kern-\belowrulesep
    \begingroup
      \color{#1}%
      \hrule height\belowrulesep
    \endgroup
  }%
}
\newcommand*{\aboverulesepcolor}[1]{%
  \noalign{%
    \begingroup
      \color{#1}%
      \hrule height\aboverulesep
    \endgroup
    \kern-\aboverulesep
  }%
}
\usepackage{amsthm,amssymb}
\usepackage{hyperref}
\usepackage{rotating}
\makeatletter
\let\@fnsymbol\@arabic
\makeatother

\newcounter{inlineenum}
\renewcommand{\theinlineenum}{(\roman{inlineenum})}
\newenvironment{inlineenum}
  {\unskip\ignorespaces\setcounter{inlineenum}{0}%
   \renewcommand{\item}{\refstepcounter{inlineenum}{\textit{\theinlineenum}~}}}
  {\ignorespacesafterend}
\usepackage{caption}
\usepackage{subcaption}

\title{\textsc{GraphTrials}: Visual Proofs of Graph Properties}

\authorrunning{F\"orster et al.}

\author{Henry F\"orster}{Department of Computer Science, University of T\"ubingen, DE}{henry.foerster@uni-tuebingen.de}{}{}

\author{Felix Klesen}{Institute of Computer Science, University of W\"urzburg, DE}{felix.klesen@uni-wuerzburg.de}{}{}
 
\author{Tim Dwyer}{Faculty of Information Technology, Monash University, AU}{tim.dwyer@monash.edu}{}{}

\author{Peter Eades}{School of Computer Science, The University of Sydney, AU}{peter.eades@sydney.edu.au}{}{}

\author{Seok-Hee Hong}{School of Computer Science, The University of Sydney, AU}{seokhee.hong@sydney.edu.au}{}{}

\author{Stephen G. Kobourov}{Department of Computer Science, University of Arizona, USA}{kobourov@cs.arizona.edu}{}{}

\author{Giuseppe Liotta}{Department of Engineering, University of Perugia, IT}{giuseppe.liotta@unipg.it}{}{}

\author{Kazuo Misue}{Department of Computer Science, University of Tsukuba, JP}{misue@cs.tsukuba.ac.jp}{}{}

\author{Fabrizio Montecchiani}{Department of Engineering, University of Perugia, IT}{fabrizio.montecchiani@unipg.it}{}{}

\author{Alexander Pastukhov}{Department of Psychology, University of Bamberg, DE}{alexander.pastukhov@uni-bamberg.de}{}{}

\author{Falk Schreiber}{Department of Computer Science, University of Konstanz, DE}{falk.schreiber@uni-konstanz.de}{}{}

\Copyright{H. F\"orster, F. Klesen, T. Dwyer, P. Eades, S.-H. Hong, S. G. Kobourov, G. Liotta, K. Misue, F. Montecchiani, A. Pastukhov and F. Schreiber}



\ArticleNo{ }

\begin{document}

\maketitle

\begin{abstract}
Graph and network visualization supports exploration, analysis and communication of relational data arising in many domains: from biological and social networks, to transportation and powergrid systems. With the arrival of AI-based question-answering tools, issues of trustworthiness and explainability of generated answers motivate a greater role for visualization. In the context of graphs, we see the
need for visualizations that can convince a critical audience that an assertion about the graph under analysis is valid. The requirements for such representations that convey precisely one specific graph property are quite different from standard network visualization criteria which optimize general aesthetics and readability. 

In this paper, we aim to provide a comprehensive introduction to visual proofs of graph properties and a foundation for further research in the area. 
We present a framework that defines what it means to visually prove a graph property. In the process, we introduce the notion of a visual certificate, that is, a specialized faithful graph visualization that leverages the viewer's perception, in particular, pre-attentive processing (e.\,g.\ via pop-out effects), to verify a given assertion about the represented graph. We also discuss the relationships between visual complexity, cognitive load and complexity theory, and propose a classification based on visual proof complexity.
Finally, we provide examples of visual certificates for problems in different visual proof complexity classes.
\end{abstract}
\acknowledgements{This work was initiated at Dagstuhl seminar 23051 ``Perception in Network Visualization''. We thank the organizers for making this fruitful interdisciplinary exchange possible and all participants for interesting discussions and insights during the seminar week.}

\begin{figure}
  \centering
  \includegraphics[width=\linewidth]{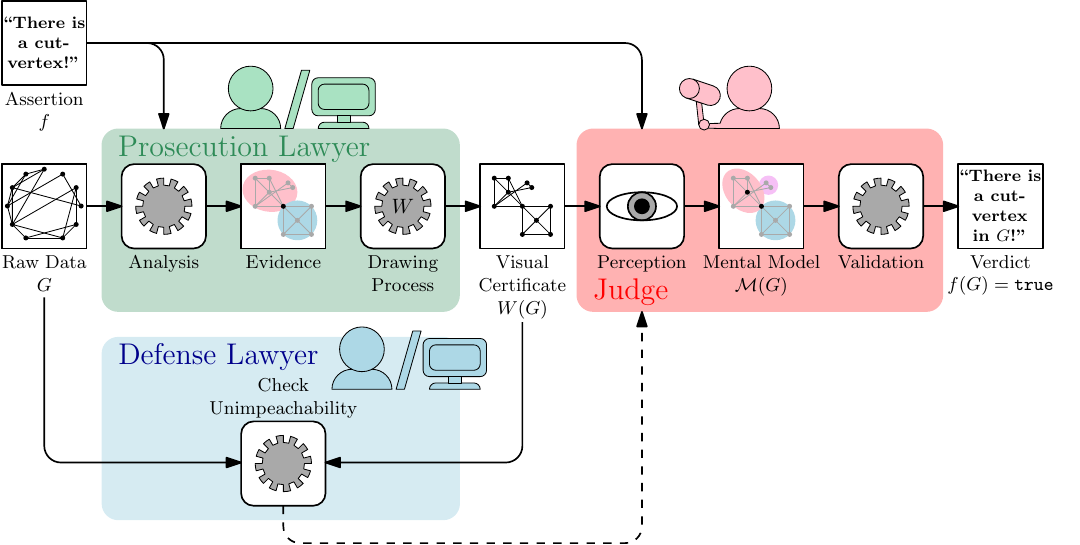}
  \vspace*{-3ex}
  \caption{Our model \textsc{GraphTrials} identifies key processes for visually proving an assertion about a given graph in an adversarial setting. The \emph{prosecution lawyer}, i.\,e., a software or a human (assisted by software), intends to highlight evidence for a graph being guilty of satisfying an assertion using a visual certificate drawing. To convince the \emph{judge}, i.\,e., the human audience of the drawing, the visual certificate guides the judge's perception to form a mental model which makes the assertion easy to validate. The visual certificate must be unimpeachable as a \emph{defense lawyer} (software or human adversary)
  checks for reasons to doubt the certificate's validity  to influence the judge's verdict.}
  \label{fig:teaser}
\end{figure}

\section{Introduction}
\label{sec:intro}
While state-of-the-art graph and network visualization techniques do a reasonable job of untangling graphs to convey meaning and support free-form exploration, there are certain application scenarios where these algorithms fall short. Namely, we focus on applications where it is necessary to \emph{convince} a (possibly non-expert) audience that a particular graph has some structural property. We emphasize that this kind of application scenario differs significantly from the traditional usage of visualization to generate new knowledge. Namely, existing graph and network visualization techniques have sought mainly  to represent all aspects of a graph or network structure as faithfully as possible such that a user can explore the visualization, identify structures, and gain insights about the underlying data. 
These traditional visualization techniques can be sufficient for journalists and other communicators to support a narrative in print or on-line media~\cite{narratingnetworks} by showing only selected views of graphs.

However, novel approaches are required in our setting in which a specific property of the data is to be conveyed in an adversarial setting where 
%
the validity of the evidence presented may be questioned (see also the defense lawyer role in Fig.~\ref{fig:teaser} which may, e.\,g., represent doubts of the audience). For example, the investigative activity of the Italian Revenue Agency (IRA) exploits the visual analysis of social networks whose nodes are the actors of potential fraudulent activities and whose edges represent financial/legal transactions between the actors. The investigators of IRA who suspect a group of persons or a single individual/company of tax evasion submit a case to the Italian financial Police for possible prosecution, which also implies showing some structural properties of the network beyond reasonable doubts. See, e.\,g., \cite{DBLP:journals/dss/DidimoGLMP18,DBLP:journals/isci/DidimoGLMP19,DBLP:journals/access/DidimoGLMMP20} for references about the use of visual analytics in the context of contrasting tax evasion in Italy.\footnote{One author has been approached by the Australian Security and Investments
Commission (a governmental regulator for stock
exchange) inquiring about visualizations to convince a court about illegal trades.} Below we describe introductory examples.

\begin{example}\label{ex:cut-vertex} A network admin discovers that two critical parts of the infrastructure would not be able to communicate with each other if a particular switch fails. To increase the robustness of the network, new hardware is needed. They have to convince the manager, who has no background in network security, to fund new hardware. \end{example}

\begin{example}\label{ex:mst}In a legal court case, the prosecution discovered that money acquired in black market sales was laundered by laundromat chain as evidenced by money provably transferred via a complicated network from the dealers to the laundromats. The prosecution has to convince the judge that all suspects belong to the criminal syndicate.\end{example}

\begin{example}\label{ex:hamiltonian}A new AI based heuristic is able to efficiently decide if a given graph is Hamiltonian, i.\,e., to test if it contains a cycle traversing all its vertices exactly once\footnote{Note that neural network approaches for NP-hard problems have been described, e.\,g., in~\cite{DBLP:conf/iclr/BelloPL0B17}. In addition, the need for visualizations in the context of explainable deep learning has been described, e.\,g., in~\cite{DBLP:journals/cga/ChooL18}.}. However, false positives must be filtered out. A human operator needs to perform this task as there is no efficient algorithm. To facilitate this, the new version of the algorithm should also create a visualization of the graph making the Hamiltonian cycle obvious to the operator.
\end{example}

Such scenarios have key differences to standard motivations for graph visualization.  Typical graph visualization techniques (node-link layout algorithms~\cite{DBLP:books/ph/BattistaETT99,DBLP:reference/crc/2013gd}, matrix ordering approaches~\cite{DBLP:journals/cgf/BehrischBRSF16} and mixed approaches which either include features of different paradigms~\cite{DBLP:journals/tvcg/HenryFM07,DBLP:conf/gd/AngoriDMPT19,DBLP:journals/tvcg/AngoriDMPT22} or show different visualizations side-by-side~\cite{DBLP:conf/vinci/BurchBCKPSV20,DBLP:journals/tvcg/HenryF06}) usually seek a representation showing as many graph properties as possible simultaneously (by trading off aesthetic and readability criteria~\cite{DBLP:journals/tvcg/AhmedLDKL22,DBLP:conf/gd/BekosFGH0SS18,DBLP:journals/cj/BekosFGHKSS21,DBLP:conf/gd/DevkotaALIK19}). However, for the scenarios above it is better to focus on showing optimally and faithfully just one specific property, i.\,e., we want a \emph{visual proof} for that property. 

More precisely, a visual proof is a proof  given by the use of a graphical or visual representation called \emph{visual certificate}. 
A good visual proof should be clear and concise, conveying the main idea in an easy-to-understand way. It should be able to effectively communicate the desired message without being overly complex or cluttered. Additionally, the visual certificate should be aesthetically pleasing and easy to interpret. Somehow, it  should be able to provide evidence to support the argument being made. Thus, a good visual certificate should be accurate, concise, and free of errors or mistakes.

In fact, visual proofs are already used in mathematics and other areas such as logic, graph theory, computer science,  and physics~\cite{nelsen1993proofs,wiki}; visual proofs are often easier to understand than algebraic proofs, as they are less abstract and easier to follow. Accessible proofs are often considered more beautiful by mathematicians; e.\,g., Appel and Haken employed  a computer-assisted proof of the long-open four-color theorem in 1976~\cite{DBLP:journals/dm/AppelH76}. This new type of proof sparked philosophical debates~\cite{philosophicalImplications} and while the theorem is  broadly accepted as proved\footnote{According to the \emph{Oxford English Dictionary}, it is yet to be proven as a ``\emph{mathematical} theorem''~\cite{oxford}.}, researchers still desire a more elegant proof~\cite{fiveColor}. Thus, we expect that visual proofs are appealing and even more convincing to experts also in fields other than mathematics.

Visual proofs can also convey properties to non-expert users or explain correctness of AI-generated solutions.
As powerful chat-based interfaces are capable of generating plausible sounding -- but difficult to verify -- explanations of complex phenomena,
we believe that there is a requirement to understand what makes a graph representation a proper visual certificate. 

\subparagraph{Contribution.} 
We introduce a model identifying important steps and their interactions in a visual proof of a graph property. Based on this model, we formalize the concept of visual certificates and give requirements for a visualization to qualify as such. We also  give examples of visual proofs for widely used graph properties and identify open research questions that should be answered to better understand visual proofs and make  them algorithmically usable.


\section{First Examples of Visual Proofs}
\label{sec:examples:easy}

\subsection{Example~\ref{ex:cut-vertex}: The Graph contains a Cut-Vertex}
\label{sec:cutvertex}

\begin{figure}[t]
    \centering
    \begin{subfigure}{0.3\linewidth}
    \centering
    \includegraphics[width=\linewidth]{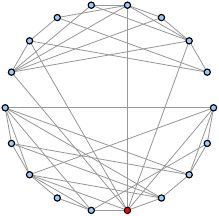}
    \subcaption{}
    \label{fig:cutvertex:circular}
    \end{subfigure}
    \hfill\begin{subfigure}{0.3\linewidth}
    \centering
    \includegraphics[width=\linewidth]{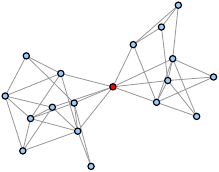}
    \subcaption{}
    \label{fig:cutvertex:force-directed}
    \end{subfigure}
    \hfill\begin{subfigure}{0.3\linewidth}
    \centering
    \includegraphics[width=\linewidth]{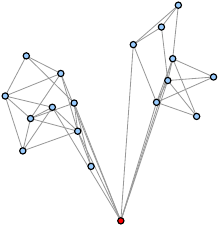}
    \subcaption{}
    \label{fig:cutvertex:proof}
    \end{subfigure}
    \caption{Three layouts generated with \textsc{yEd}\footnotemark~\cite{yed}. (a) and (b)~Circular and organic layouts generated with standard settings, resp. (c)~Manually created layout highlighting the cut-vertex.}
    \label{fig:cutvertex}
\end{figure}

First, we revisit the situation in Ex.~\ref{ex:cut-vertex}. In this communication network there are two distinct parts such that all connections between them traverse a single switch. 
This corresponds to the graph underlying the network containing a \emph{cut-vertex}, whose removal separates the remainder of the graph into at least two distinct components.
Hence, in order to convince the manager, the network admin has to point out that the graph underlying the network can be separated by the removal of  the vertex corresponding to the switch. So, they first layout the graph using a circular layout, which is a wide-spread all-purpose visualization style~\cite{ST99}, and point the manager to the fact that the red colored vertex is a cut-vertex; see \cref{fig:cutvertex:circular}.  

Unfortunately, the circular layout does a poor job at highlighting the cut-vertex. While it is evident to the manager that there are a top and a bottom component connected by some edges, they explain that they are not sure if all connections between both components use the suggested cut-vertex or not. Hence, the network admin prepares a second drawing using a force-directed organic layout where the cut-vertex is clearly visible; see~\cref{fig:cutvertex:force-directed}. 
However, the engineer  who designed the network becomes defensive and claims that there could be another edge hidden behind the alleged cut-vertex. This argument can be easily disproven by the network admin as they move the cut-vertex down, obtaining the drawing in~\cref{fig:cutvertex:proof}. Presented with this new line of evidence, the engineer stops arguing and the manager agrees that the network has to be made more robust.
\footnotetext{Unless specified otherwise,  the layouts of all visualizations in this paper have been created by the authors.}

\subparagraph{Discussion.} This example illustrates  how standard layout techniques may be unable to highlight even simple properties. 
In the circular layout, it is not easy to verify  even when the cut-vertex is highlighted; see~\cref{fig:cutvertex:circular}. 
This is due to the  Gestalt principle of grouping~\cite{Wagemans2012,Ware}. Here, the initial perception is guided by continuity and closure of node positions, leading to the perception of a single circular component.  As a second step, an observer may see two separate components with edges biasing perception due to connectedness grouping. Thus, the observer has to analyze the entire graph, going node-by-node, to negate the automatic perceptual grouping induced by the layout to verify that there is a cut-vertex. 
The issue with the second illustration in~\cref{fig:cutvertex:force-directed} is of different nature. Namely, the force-directed layout does a much better job at highlighting the cut-vertex. In fact, the observer discovers two dense salient features which are the two components separated by the cut-vertex and immediately notes that they are connected at a single vertex. 
Nevertheless, if there is an  overlapping edge behind the cut-vertex, the drawing may look the same, challenging the human observer to identify that the vertex is not a cut-vertex.
%
The drawing in~\cref{fig:cutvertex:proof} avoids this problem by explicitly highlighting the cut-vertex via pre-attentively perceptable patterns (i.\,e., pop-out effects)~\cite{Ware}. The singular goal of highlighting the cut-vertex is achieved at the cost of traditionally accepted aesthetic metrics~\cite{Purchase2002}, as --- compared to the circular and force-directed layouts --- the general layout is unbalanced, with many crossings and poor resolution; see Table~\ref{tab:metrics}.  
Thus, visual certificates may not  be useful in traditional exploratory applications, instead they focus on highlighting a specific property.

\begin{figure}[t]
    \centering
    \begin{subfigure}{0.3\linewidth}
    \centering
    \includegraphics[width=\linewidth]{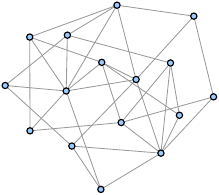}
    \subcaption{}
    \label{fig:st:bad}
    \end{subfigure}
    \hfill\begin{subfigure}{0.3\linewidth}
    \centering
    \includegraphics[width=\linewidth]{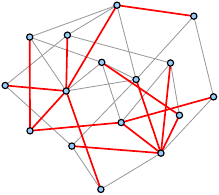}
    \subcaption{}
    \label{fig:st:middle}
    \end{subfigure}
    \hfill\begin{subfigure}{0.3\linewidth}
    \centering
    \includegraphics[width=\linewidth]{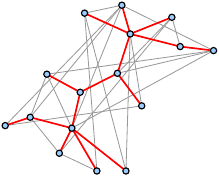}
    \subcaption{}
    \label{fig:st:good}
    \end{subfigure}
    \caption{(a)--(b)~Organic layout generated with standard settings by \textsc{yEd}~\cite{yed}, with a  spanning tree highlighted in (b). (c)~A manually created layout highlighting the spanning tree.}
    \label{fig:st}
\end{figure}

We remark that a cut-vertex proves non-$2$-connectivity and a similar approach can be used 
to visually prove  that a graph is \emph{not $k$-connected}: there exists a set of $k-1$ vertices whose removal separates the graph and we can layout the graph so that all connections between two clearly separated parts run via this vertex set. 

\subsection{Example~\ref{ex:mst}: The Graph is Connected}
\label{sec:connected}

In Ex.~\ref{ex:mst}, to convince the judge, the prosecution lawyer decides to visualize the network of criminals induced by the connections of provable money transfers. The prosecution lawyer draws it with a force-directed approach; see~\cref{fig:st:bad}. 
While \cref{fig:st:bad} shows that there are many connections in the graph, it does not emphasize that there is only a single connected component. Hence, the defense lawyer argues that the component containing their client may have been drawn on top of the component with all the convicted criminals. Hence, the prosecution lawyer has to  improve their visual proof. To do so, they include a highlighted \emph{spanning tree} that shows that every vertex can be reached from every other vertex; see~\cref{fig:st:middle}. %
Although the defense lawyer now has to admit that there is a smaller portion of the drawing to check, i.\,e., the highlighted edges, their argument stays more or less the same: that there are still crossings between edges of the spanning tree, which may be due to two different highlighted components drawn on top of each other. Thus, the prosecution lawyer creates a third drawing in which the spanning tree is crossing-free; see~\cref{fig:st:good}. Here the spanning tree is rooted at the central vertex and vertices are drawn on concentric circles depending on distance from the root. 
Given this visualization, the defense rests, and the judge decides quickly that indeed all members of the network are affiliated.

\setlength\extrarowheight{1pt}
\begin{sidewaystable*}[p]
\caption{Selected aesthetic metrics of the node-link drawings in this paper: stress \textsc{st}~\cite{DBLP:journals/ipl/KamadaK89},  node resolution \textsc{nr}, Jaccard index \textsc{ji}~\cite{DBLP:journals/tvcg/AhmedLDKL22}, edge-length ratio \textsc{el}~\cite{DBLP:conf/gd/LazardLL17,DBLP:journals/tcs/LazardLL19},  crossing resolution \textsc{cr}~\cite{DBLP:conf/gd/GiacomoDLM09,DBLP:journals/mst/GiacomoDLM11},  crossing number \textsc{cn}, aspect ratio \textsc{ar} and angular resolution \textsc{an}~\cite{DBLP:books/ws/NishizekiR04}. Red numbers in parantheses give the corresponding values for subgraphs highlighted in red in the corresponding figure. For \textsc{st} and \textsc{cn} lower numbers are better, otherwise higher numbers are better.}
\small \centering \begin{tabular}{
>{\columncolor[HTML]{A2F5A2}}c ccc
>{\columncolor[HTML]{E2E2E2}}c 
>{\columncolor[HTML]{E2E2E2}}c 
>{\columncolor[HTML]{E2E2E2}}c cc
>{\columncolor[HTML]{E2E2E2}}c ccc
>{\columncolor[HTML]{E2E2E2}}c 
>{\columncolor[HTML]{E2E2E2}}c }
\hline
\cellcolor{blue!33}                                & \multicolumn{3}{c}{\cellcolor{blue!25}   \ref{fig:cutvertex}}                                           & \multicolumn{3}{c}{\cellcolor{blue!33}   \ref{fig:st}}                                           & \multicolumn{2}{c}{\cellcolor{blue!25}   \ref{fig:hamilton}}             & \cellcolor{blue!33}   \ref{fig:coloring}  & \multicolumn{3}{c}{\cellcolor{blue!25}   \ref{fig:nonbipartite}}                                           & \multicolumn{2}{c}{\cellcolor{blue!33}   \ref{fig:noncomplete}}             \\
\multirow{-2}{*}{\cellcolor{blue!33}   \textbf{Fig.}} & \cellcolor[HTML]{C0C0C0}\hyperref[fig:cutvertex:circular]{(a)} & \cellcolor[HTML]{C0C0C0}\hyperref[fig:cutvertex:force-directed]{(b)} & \cellcolor[HTML]{C0C0C0}\hyperref[fig:cutvertex:proof]{(c)}& \multicolumn{2}{c}{\cellcolor[HTML]{B0B0B0}\hyperref[fig:st:bad]{(a)} -- \hyperref[fig:st:middle]{(b)}} &\cellcolor[HTML]{B0B0B0}\hyperref[fig:st:good]{(c)} & \cellcolor[HTML]{C0C0C0}\hyperref[fig:hamilton1]{(a)} & \cellcolor[HTML]{C0C0C0}\hyperref[fig:hamilton2]{(b)} & \cellcolor[HTML]{B0B0B0}\hyperref[fig:coloring4]{(d)} & \cellcolor[HTML]{C0C0C0}\hyperref[fig:nonbipartite0]{(a)} & \cellcolor[HTML]{C0C0C0}\hyperref[fig:nonbipartite2]{(b)} & \cellcolor[HTML]{C0C0C0}\hyperref[fig:nonbipartite3]{(c)} & \cellcolor[HTML]{B0B0B0}\hyperref[fig:noncomplete1]{(a)} & \cellcolor[HTML]{B0B0B0}\hyperref[fig:noncomplete2]{(b)} \\ \hline
\cellcolor[HTML]{F7A6A6}\textsc{st}                              & 132.1                       & 9.2                         & 36.7                        & \multicolumn{2}{c}{\cellcolor[HTML]{E2E2E2}13.6 (\textcolor{red}{31.3})}  & 32.5 (\textcolor{red}{19.6})                     & 9.3 (\textcolor{red}{51.9})                  & 45.5 (\textcolor{red}{12.6})                 & 6635                        & 567.6 (\textcolor{red}{9.1})                 & 651.0 (\textcolor{red}{1.9})                 & 1022.7 (\textcolor{red}{3.3})                & 58.8                        & 94.4                       \\
\cellcolor[HTML]{F7A6A6}\textsc{cn}                              & 81                          & 19                          & 26                          & \multicolumn{2}{c}{\cellcolor[HTML]{E2E2E2}28 (\textcolor{red}{8})}       & 63 (\textcolor{red}{0})                          & 29 (\textcolor{red}{4})                      & 45 (\textcolor{red}{0})                      & 7452                        & 6211 (\textcolor{red}{2})                    & 9493 (\textcolor{red}{0})                    & 9992 (\textcolor{red}{0})                    & 12649                       & 12650                      \\
\textsc{ji}                                                      & .274                        & .332                        & .283                        & \multicolumn{2}{c}{\cellcolor[HTML]{E2E2E2}.330 (\textcolor{red}{.132})}  & .285 (\textcolor{red}{.160})                     & .407 (\textcolor{red}{.167})                 & .339 (\textcolor{red}{.232})                 & .037                        & .177 (\textcolor{red}{.243})                 & .180 (\textcolor{red}{.361})                 & .182 (\textcolor{red}{.361})                 & .918                        & .920                        \\
\textsc{el}                                                     & .174                        & .333                        & .127                        & \multicolumn{2}{c}{\cellcolor[HTML]{E2E2E2}.360 (\textcolor{red}{.404})}  & .149 (\textcolor{red}{.382})                     & .401 (\textcolor{red}{.649})                 & .185 (\textcolor{red}{.542})                 & .090                        & .020 (\textcolor{red}{.161})                 & .023 (\textcolor{red}{.586})                 & .013 (\textcolor{red}{.472})                 & .126                        & .047                        \\
\textsc{nr}                                                     & .174                        & .104                        & .100                        & \multicolumn{2}{c}{\cellcolor[HTML]{E2E2E2}.181 (\textcolor{red}{.181})}         & .119    (\textcolor{red}{.119})                        & .164 (\textcolor{red}{.164})                        & .184      (\textcolor{red}{.184})                  & .020                        & .016 (\textcolor{red}{.123})                 & .016 (\textcolor{red}{.314})                 & .011 (\textcolor{red}{.307})                 & .126                        & .047                        \\
\textsc{ar}                                                    & .985                        & .774                        & .966                        & \multicolumn{2}{c}{\cellcolor[HTML]{E2E2E2}.884 (\textcolor{red}{.884})}         & .796    (\textcolor{red}{.796})                        & .871     (\textcolor{red}{.871})                   & .987     (\textcolor{red}{.987})                   & 1                           & .941 (\textcolor{red}{.606})                 & .941 (\textcolor{red}{.977})                 & .994 (\textcolor{red}{.994})                 & .998                        & .898                        \\
\textsc{cr}                                                    & 20.0                        & 27.1                        & 7.2                         & \multicolumn{2}{c}{\cellcolor[HTML]{E2E2E2}37.1 (\textcolor{red}{37.1})}         & 20.5 (\textcolor{red}{N/A})                       & 31.4 (\textcolor{red}{48.2})                 & 22.5 (\textcolor{red}{N/A})                   & 2.6                         & 0.79 (\textcolor{red}{73.2})                 & 0.79 (\textcolor{red}{N/A})                   & 0.74 (\textcolor{red}{N/A})                   & 14.4                        & 4.68                        \\
\textsc{an}                                                     & 10.0                        & .20                         & .56                         & \multicolumn{2}{c}{\cellcolor[HTML]{E2E2E2}1.68 (\textcolor{red}{20.0})} & 4.44 (\textcolor{red}{21.4})                     & 0.40 (\textcolor{red}{15.93})                & 12.08 (\textcolor{red}{141.0})               & 0.21                        & .017 (\textcolor{red}{6.87})                 & .017 (\textcolor{red}{112.0})                 & .007 (\textcolor{red}{97.6})                 & 7.20                        & 1.23                        \\ \hline
\end{tabular}
\label{tab:metrics}
\end{sidewaystable*}
\setlength\extrarowheight{0pt}

\subparagraph{Discussion.} While in Ex.~\ref{ex:cut-vertex} we have seen that the drawing style of the entire graph can be important to visually prove a property, here we added another dimension. Namely, a subgraph is explicitly color-highlighted for pre-attentive perception. 
%
In addition, the drawing of this subgraph was very important in creating a convincing argument. In~\cref{fig:st:middle} the drawing of the spanning tree is not very readable. Thus, even with the attention drawn to this portion of the drawing, it remains time consuming to check that a single tree connects all vertices. But when the tree is laid out in a concise and readable fashion as in~\cref{fig:st:good}, it is quite evident that it spans all the vertices, as the colored edges induce automatic grouping via similarity~\cite{Wagemans2012} and act as guidance for attention spread~\cite{Houtkamp2003}. Similar to \cref{ex:cut-vertex}, while the quality of the drawing of the spanning tree is improved, the drawing of the rest of the graph does not measure well on the usual metrics; see \cref{tab:metrics}.  

\begin{figure}[t]
    \centering
    \begin{subfigure}{0.225\linewidth}
    \centering
    \includegraphics[width=\linewidth]{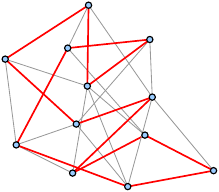}
    \subcaption{}
    \label{fig:hamilton1}
    \end{subfigure}
    \hfill\begin{subfigure}{0.225\linewidth}
    \centering
    \includegraphics[width=\linewidth]
    {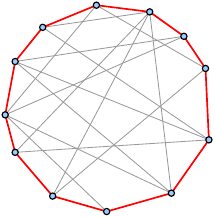}
    \subcaption{}
    \label{fig:hamilton2}
    \end{subfigure}
    \hfill\begin{subfigure}{0.225\linewidth}
    \centering
    \includegraphics[width=\linewidth]
    {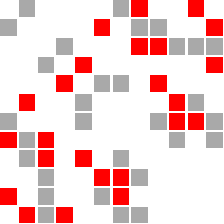}
    \subcaption{}
    \label{fig:hamilton4}
    \end{subfigure}
    \hfill\begin{subfigure}{0.225\linewidth}
    \centering
    \includegraphics[width=\linewidth]
    {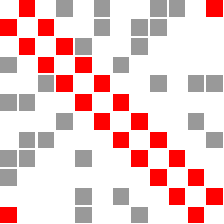}
    \subcaption{}
    \label{fig:hamilton3}
    \end{subfigure}
    \caption{Four layouts of a graph with a Hamiltonian cycle (red).}
    \label{fig:hamilton}
\end{figure}

\subsection{Example~\ref{ex:hamiltonian}: The Graph has a Hamiltonian Cycle}
\label{sec:hamilton}

We may train an AI to produce a good node-link drawing for \cref{ex:hamiltonian}. As in~\cref{sec:connected}, we observe that the quality of the drawing of the evidence relevant to the property under consideration is more important than the drawing of the full graph (\cref{fig:hamilton1}), thus we select a circular layout with the hamiltonian cycle forming the outer face (\cref{fig:hamilton2}). 
However, the human operator needs to check that all edges of the highlighted outer cycle are indeed present which can become increasingly difficult for larger graphs where resolution may become problematic. We can improve upon these issues by instead using an adjacency matrix representation. While an arbitrary permutation (see~\cref{fig:hamilton4}) does not provide any insights, an appropriate sorting of rows and columns makes the cycle composed of three components: one red diagonal and two red-cells (top-right and bottom-left); see~\cref{fig:hamilton3}.

\subparagraph{Discussion.} We observed that different visualization paradigms may perform better or worse for visually proving a property. While the node-link drawing in~\cref{fig:hamilton2} already highlights the cycle well, the adjacency matrix representation in~\cref{fig:hamilton3} composes the Hamilton cycle in three components. The perception of the red diagonal is facilitated by figure-ground separation via connectedness and similarity~\cite{Wagemans2012}, and  the two corner cells stand out due to both color difference and symmetry~\cite{Wolfe2010}.  The particular advantage of this representation is that the used visual cues scale nicely even for very large matrices (up to the pixel resolution of the screen)~\cite{Wolfe2010}. Thus, an important criterion for judging the quality of a visual proof should be the workload required by the observer to evaluate the correctness. As checking for Hamiltonicity is a difficult task with an all-purpose visualization (see also~\cref{sec:complexity}), both visualizations should be regarded as valid visual certificates albeit of different quality.

\section{Related Theories, Frameworks and Models}
\label{sec:preliminaries}


\subsection{Certifying Algorithms}
\label{sec:certiying-algorithms}
The concept of visual certificates is related to \emph{certifying algorithms} popularized by McConnell et al.~\cite{DBLP:journals/csr/McConnellMNS11}, which seek to provide short and easy-to-check \emph{certificates} for the correctness of an algorithm. 
%
%
%
Let $f:X \rightarrow Y$ be a computable, surjective function for input set $X$ and output set $Y$ and let $W$ be a set of \emph{witnesses}. Intuitively speaking, a witness describes a simple proof certifying that the output $y$ of an algorithm for $f$ on input $x$ satisfies $f(x)=y$. The validity of a witness for a certain combination of inputs and outputs is assessed via the witness predicate $\mathcal{W}: X \times Y \times W \rightarrow \{\texttt{true},\texttt{false}\}$ that fulfills:

\begin{enumerate}
    \itemsep0em 
    \item\label{prop:certifying:1} \emph{Witness property:} Given $(x,y,w) \in X \times Y \times W$, it holds $f(x)=y \Leftrightarrow \mathcal{W}(x,y,w) = \texttt{true}$.
    \item\label{prop:certifying:2} \emph{Checkability:} Given $(x,y,w) \in X \times Y \times W$, it is trivial to determine $\mathcal{W}(x,y,w)$. 
    \item\label{prop:certifying:3} \emph{Simplicity:} $\mathcal{W}(x,y,w)  \Rightarrow  f(x)=y$ has a simple proof.
\end{enumerate}

An algorithm for $f$ is now called \emph{certifying algorithm} if for any input $x \in X$ it computes the output $y=f(x) \in Y$ and a witness $w \in W$ such that $\mathcal{W}(x,y,w) = \texttt{true}$.
It is worth noting that Properties~\ref{prop:certifying:2} and~\ref{prop:certifying:3} of the witness predicate are vaguely formulated. McConnell et al.~\cite{DBLP:journals/csr/McConnellMNS11} suggest that Property~\ref{prop:certifying:2} can be formalized by requiring that there must be a decision algorithm for $\mathcal{W}$ that runs in a certain time (such an algorithm is  called a \emph{checker}). On the other hand, they emphasize that Property~\ref{prop:certifying:3} is intentionally left subjective as it relies on what is 
considered common knowledge. For examples of certifying algorithms, see \cref{app:certifying}.

\subsection{Perception}
Visual proofs are concerned with a design of visual evidence for an existence of a specific property, such as the presence of a cut-vertex, a Hamiltonian cycle, etc.  In principle, a proof for such a property can be reduced to a program that returns a binary outcome, affirming or rejecting the claim.  This may be sufficient for specialists who are familiar with the property itself, understand and trust the algorithm behind the code, and trust that the code is valid. However, such evidence may not be convincing to a non-specialist (a judge, a stockholder, etc.), particularly because the proof itself will be just one piece of evidence among many. Prior research shows that in such cases, presenting evidence \textit{per se} is not enough, as information can be discounted as confusing, unimportant, or, given the wrong context, even misleading~\cite{Tufte2001}, as the accessibility and clarity of evidence could be as important as evidence itself~\cite{Healy2018}.

Due to the diversity of graph properties there can be no general solution. Visual proof design might be guided by the principle of optimizing the \emph{data-ink} ratio~\cite{Tufte2001}. Thus, instead of optimizing overall aesthetic metrics~\cite{Purchase2002}, one should minimize the required number of \emph{visual queries}, i.\,e., attention orientation, driving eye movements, and pattern/object recognition~\cite{Ware2021}.

The human visual perception system consists of three stages: (1) rapid parallel processing involving billions of neurons, e.\,g.,  extraction of orientation, texture, color, and motion features; (2) slower processing than Stage 1, e.\,g.,  detection of 2D patterns, contours and regions; (3) slow serial processing, involving both working and long-term memory, e.\,g.,  object identification~\cite{Ware}.
As in Stage 1 the entire visual field is processed quickly in parallel, information that can be captured in this stage can be easily distinguished. 
Thus, pre-attentive (pop-out) patterns such as color, size, orientation, shapes, etc.\  
should be utilized.

In other words, a good visual proof must ensure that a focal piece of evidence is a visual “pop-out” feature that automatically attracts viewer attention and that the visual layout is 
parsed and grouped into patterns that express the evidence. In case of the former, studies on visual search provide a comprehensive list of useful pop-out features such as color, size, contrast, or location~\cite{Wolfe2010}. Regarding the latter, one can rely on a large body of literature on principles of perceptual organization, commonly known as Gestalt principles~\cite{Wagemans2012}. However, yet another constraint is placed by our working memory that limits the number of nodes, edges, and components that can realistically be assessed at any single time~\cite{Luck2008}.

The examples  above illustrate the importance of this approach for visual proofs. For instance, consider the visual evidence for the existence of a cut-vertex in \cref{fig:cutvertex:circular}.  While it uses color to attract the viewer’s attention to the cut-vertex and spatial arrangement to visually separate the two components, it still leads to an excessive number of visual queries, requiring multiple scans of individual vertices to ensure that they are connected only to the cut-vertex and the nodes within the component. In turn,  there is a memory bottleneck that is likely to prevent a viewer from  being completely certain about the validity of the proof. In contrast, in \cref{fig:cutvertex:proof} the graph layout groups the entire evidence in just three components and clearly shows lack of inter-component edges, so that very few visual queries are required to confirm the vertex is indeed a sole connector between the components. 
In short, although there cannot be a single one-size-fits-all approach for constructing visual proofs, their critical role in aiding the cognition of the viewer means they should be built based on principles of perceptual organization and around the limitations of attention and memory~\cite{Ware2021}.

\subsection{Computational Complexity}
\label{sec:complexity}
To evaluate the amount of the cognitive workload, we will apply concepts from complexity theory~\cite{conp,DBLP:books/fm/GareyJ79}.
It is also worth mentioning that the examples discussed so far differ in terms of their computational complexity. Namely, all cut-vertices of a graph and a spanning-tree can be found in time $O(n+m)$ based on BFS traversals where $n$ is the number of vertices and $m$ the number of edges while determining a Hamiltonian cycle is  \emph{NP-complete}~\cite{DBLP:books/fm/GareyJ79}. Thus, in~\cref{ex:hamiltonian}, we have visually proven an algorithmically difficult to solve problem. 




However, there may be graph properties that cannot be visually proven.
We first have to discuss how a human observer interacts with a visual certificate. In~\cref{ex:cut-vertex}, the human observer identified two connected components and then saw that they can be separated by the removal of their shared vertex. Such a procedure could be seen as an $O(1)$ time algorithm, where the observer determined that there is only a single point where both components touch. Similarly, in Examples~\ref{ex:mst} and \ref{ex:hamiltonian}, the observer may have checked for every vertex if it was part of the highlighted structure. Even if they were to check this for every vertex one at a time, the resulting algorithm would still run in linear time. Hence, 
an observer is actually performing a \emph{deterministic} validation algorithm for establishing that a certificate is correct.

Now, consider the complementary question to~\cref{ex:hamiltonian}, i.\,e., we want to determine whether a graph does \emph{not} contain a Hamiltonian cycle. This is a \emph{CoNP-complete} problem as it is the complement to an {NP-complete problem}. For CoNP-complete problems it is likely that there is no certificate that can be checked in polynomial time~\cite{conp}, i.\,e., if we assume that a human observer deterministically analyzes a visualization (as could be recreated by computer vision), we have to assume that we cannot visually prove a CoNP-complete problem.

\subsection{Related Visualization Models}

%
Aside from graph visualizations, the concept of visually enhancing a proof is wide-spread. In mathematics, visual proofs for theorems have been used since ancient times~\cite{wiki} and there is a plethora of examples~\cite{nelsen1993proofs}. The question if such proofs can be regarded as such also has been discussed philosophically~\cite{philosophy}. Also in computer science, visualizations are heavily used to convey knowledge, e.\,g., while not necessarily proving, an interactive sequential art by Bret Victor~\cite{art} beautifully explained an algorithm from a Nature paper~\cite{nature}.

Overall, there is a trend of increasingly sophisticated models considering an holistic integration of visualization into the sensemaking process, typically with the goal of informing the design of interactive systems for data exploration.  Early models considered a linear pipeline, from data, via various transformations, to a visual display \cite{card1999readings}.  Visual analytics seeks to apply visualization to support the entire human sense-making loop \cite{pirolli2005sensemaking}.  More recent models aim to connect sense-making from interactive data visualization, via hypothesis formation and testing, to knowledge generation \cite{knowledgegen}.  An underlying theme across most of this work is the role of computational guidance in the analytics process, and how algorithms can support the various loops in the sensemaking process \cite{ceneda2019review}. By contrast, we consider a different model  to conceptualize the role of algorithms, and AI, in supporting data (specifically network data) understanding. Our model for visual proofs (Fig.~\ref{fig:teaser}) does not seek to replace the traditional sense-making/knowledge-generation loop, but to support humans in situations where the result of a complex algorithm or  property needs to be explained and justified.

There are also  models related to ours from information visualisation research.
Song et al.~\cite{9966829} considered a problem that may be seen as a complementary question to the one studied in this paper: They investigated how computer vision can understand network visualizations optimized for human users.
Wickham et al.~\cite{DBLP:journals/tvcg/WickhamCHB10} proposed a two-phase procedure to convince a human observer that a data set contains statistically significant difference from randomly generated data. 
The human observer is first exposed to several randomly generated data sets (similar to a Rorschach test) before being exposed to a line-up consisting of the real data set and a couple randomly generated data sets. The first phase primes the human viewer for statistically insignificant variations so that, in the second phase,  statistically significant differences clearly pop out from the noise. 
Another related model are \emph{Gragnostics}, which are ten features suggested by Gove~\cite{DBLP:conf/iv/Gove19,Gove2022}, that are fast to compute and provide a quantification of structural graph properties. In contrast to our model that aims to prove structural properties of graphs, Gragnostics provides the human user with a first impression of the structure of the graph at hand which may be helpful for initiating a thorough investigation. Finally, our model may also be seen as a visual communication of structural graph properties. Visual communication has been investigated in other settings for several decades, see e.\,g.~\cite{DBLP:books/daglib/0001351,DBLP:books/lib/Tufte97}.

\section{The \textsc{GraphTrials} Model}
\label{sec:model}

We are now ready to discuss our formalization of \emph{visual proofs}. For this, we first abstractly outline the process of visually proving properties of graphs in an adversarial setting using a model that we call \textsc{GraphTrials}; see also~\cref{fig:teaser}. The model includes three distinct roles that have already appeared in our discussion of~\cref{ex:mst} in~\cref{sec:connected}:
The \emph{prosecution lawyer} must convince the judge that a  certain assertion regarding a graph is true, the \emph{defense lawyer} may raise doubts about the validity of the prosecution lawyer’s claims, and the \emph{judge} will determine the truth of the assertion.
%
%
The roles \emph{prosecution lawyer}, \emph{judge} and \emph{defense lawyer} are to be seen as abstract descriptions of the different actors in the process; e.\,g., in Ex.~\ref{ex:cut-vertex} and~\ref{ex:hamiltonian}, the prosecution lawyers were the network admin and the AI based algorithm, respectively. The latter example further indicates that not all roles have to be assigned to a human. In fact, we only require that the judge corresponds to the human audience of the visual certificate whereas each lawyer may be either human, software or a human assisted by software. Moreover, as we have seen in~\cref{sec:cutvertex}, it can also occur that a critical audience can act as both the {judge} and {defense lawyer} roles simultaneously.

To convince the judge of a valid assertion $f$ for the input graph $G$, the \emph{prosecution lawyer} draws a visual certificate $W(G)$. To do so, they first analyze the raw data $G$ to reveal evidence that proves the assertion $f$.
The evidence is then embedded in $W(G)$: a visual representation of $G$ that in some way emphasizes the evidence. 
Note that in the scope of our model we treat the analysis of the raw data and extraction of the evidence as a black box, i.\,e., we may assume that the prosecution lawyer already knows that the assertion $f$ is true for the input graph $G$ and may also be given the evidence as input. This allows us to \emph{efficiently} visually prove algorithmically difficult assertions (such as the existence of a Hamiltonian cycle as in~\cref{sec:hamilton}) and to ignore how the evidence is gathered (either algorithmically or by human interaction) in our model. The latter aspect also provides the possibility to separate the evidence gathering from the visualization process $W$, i.\,e., $W$ could be a reusable program that embeds the evidence according to a specification\footnote{The examples in~\cref{sec:hamilton} and~\ref{sec:examples:advanced} both use visual certificates that highlight cycles.}.

The \emph{defense lawyer} checks the {\em unimpeachability} of $W(G)$ as a visual representation of $G$ certifying $f(G)$. Thus, they may question whether the graph represented in the visualization actually corresponds to the input and they may also raise concerns if $W(G)$ is not distinguishable from a slightly different non-certificate (e.\,g., in~\cref{sec:cutvertex} we encountered the case where an edge may have been hidden making it invisible to the judge's perception). 

The \emph{judge}, the human audience of the visual certificate $W(G)$, will validate the claim $f(G)$ using $W(G)$. In this step, the visual certificate $W(G)$ must guide the judge's perception so that they are able to form a mental model $\mathcal{M}(G)$ that facilitates confirmation of the validity of the assertion $f(G)$. For instance, the guidance can be formed by a suitable choice of topology  which leads the judge to identify clusters of the graph as distinct salient features (as in~\cref{sec:cutvertex}) or by adding additional features such as color to draw attention to certain parts of the graph (as in~\cref{sec:connected}). We discuss the judges mental model in~\cref{sec:mental-model}.

It is noteworthy that aside from the input graph and the verdict of the judge, the only information shared by all three roles is the \emph{visual certificate} $W(G)$. In particular, it is the only medium that can be used by the prosecution lawyer to communicate the gathered evidence to the judge, i.\,e., the evidence is hidden information only accessible by the prosecution lawyer. Similarly, the judge is not communicating its mental model $\mathcal{M}(G)$ to the prosecution or defense lawyer, yet as we discussed above both roles might want to \emph{estimate} what the mental model will look like. Furthermore, the nature of the mental model plays an important role in the validation step performed by the judge. Namely, the cognitive load put on the judge in this step depends hugely on how \emph{complex} $\mathcal{M}(G)$ is. 
Finally, the defense lawyer's checking for unimpeachability is a process that is independent of the judge and prosecution lawyer and for a \emph{faithful and readable} visual certificate we demand that there is no reason for the defense lawyer to raise doubts to the judge. As a result, there are several properties that we require from a visualization in order to call it a visual certificate and it could occur that an assertion cannot be \emph{visually proven} for every graph for which the assertion is true (for instance we discussed issues related to scalability in~\cref{sec:hamilton}). To this end, we also state when we want to say that a certain assertion can be visually proven for arbitrary graphs.

\subsection{Visual Certificates and Visual Provability}
\label{sec:certificate}

We  give formal requirements 
inspired by the concept of certifying algorithms discussed in~\cref{sec:certiying-algorithms}.
Let $f: \mathcal{G} \rightarrow \{\texttt{true},\texttt{false}\}$ be an \emph{assertion function} for the set of graphs $\mathcal{G}$, i.\,e., for some graphs the assertion $f(G)$ is  \texttt{true} while for others it is not. For instance, if $f$ is the existence of a cut-vertex, some graphs do contain one ($f(G)=\texttt{true}$) while others do not ($f(G)=\texttt{false}$). Consider a graph $G$ with $f(G)=\texttt{true}$ and let $W(G)$ be a visualization of $G$. We call $W(G)$  \emph{visual certificate} for $f(G)$ if and only if the following hold:
\begin{enumerate}
    
    \itemsep0em 
    \item \label{prop:certificate:1}\emph{Unimpeachability:} 
    We call $W(G)$ \emph{unimpeachable}, if it satisfies the following two properties. First, $W(G)$ should provide \emph{information faithfulness}~\cite{faithful}, i.\,e., it displays the ground truth properties and structures in $G$. Second,  $W(G)$ should provide \emph{task readablility}~\cite{faithful}, i.\,e., the judge can \emph{perceive} enough information for validating the assertion.  

    \item \label{prop:certificate:2}\emph{Checkability:} Given $W(G)$, it is trivial to decide that $f(G)= \texttt{true}$. In particular, this means that the judge's perception leads to the formation of a \emph{mental model} $\mathcal{M}(G)$ that makes it possible for the judge to \emph{efficiently}  validate the assertion. The number of distinct observations made by the  judge in the process is called the \emph{perceptual complexity}.   
    \item \label{prop:certificate:3}\emph{Simplicity:} Given $\mathcal{M}(G)$, there is a \emph{simple formal proof} for $f(G)= \texttt{true}$ that relies solely on conclusions that the judge may deduce using $\mathcal{M}(G)$. In particular, this means that $W(G)$ is \emph{perceptually distinguishable} from any possible wrong visual certificate $W'(G)$.
\end{enumerate}

If a visual certificate $W(G)$ exists for each $G \in \mathcal{G}$ with $f(G)=\texttt{true}$, we call $f$ \emph{visually provable}. Note that the complementary function $f^c$ (which is \texttt{true} if and only if $f(G)=\texttt{false}$) needs not  necessarily be visually provable. For instance, we were able to visually prove the assertion that $G$ contains a Hamiltonian cycle in~\cref{sec:hamilton} but we argued that the absence of such a cycle cannot be visually proven in~\cref{sec:complexity}. This and requiring unimpeachability are clear differences to the concept of certifying algorithms  whereas checkability and simplicity occur in both models, here considering the  perceptual abilities of the judge; see also~\cref{sec:certiying-algorithms}.

We are also interested in \emph{how efficiently} the judge is able to validate $f(G) = \texttt{true}$ based on $\mathcal{M}(G)$.  To this end, we define the \emph{perceptual complexity} as the time that the judge needs to check the assertion given $\mathcal{M}(G)$.  The perceptual complexity may depend on the size of the graph, however, in some scenarios (e.g. \cref{ex:cut-vertex}) it may be independent of it. Since we assume the judge to make an objective judgment based on the evidence, we can treat the thought process as a deterministic algorithm and apply methods from complexity theory to evaluate the perceptual complexity.
See \cref{app:certificate} for an application of these concepts.

\section{Visual Proofs for Graph Properties}
\label{sec:examples:advanced}

We provide  visual proofs for further widely used assertions. 
For a summary of our discussion, refer to~\cref{tab:overview}. In addition, we discuss further assertions in \cref{app:more-examples}. 

\begin{table}[t]
    \caption{Visual proofs  in this paper,  computational complexity of the problem and perceptual complexity of presented visual proofs ($n$ and $m$ denote the numbers of vertices and edges, resp.).} 
    \centering
    \begin{tabular}{cccc}
        \toprule
         \cellcolor{blue!25} \bf Assertion & {\cellcolor{blue!25}\bf Comp. Complexity} & {\cellcolor{blue!25}\bf Percep. Complexity} & \cellcolor{blue!25} \bf Sec.\\
          \aboverulesepcolor{blue!25}
         \midrule
         $G$ is connected  & \multirow{2}{*}{$O(n$+$m)$} & $O(n)$ & \ref{sec:connected} \\
         $G$ is \emph{not} (2-)connected & & $O(1)$ & \ref{sec:cutvertex} \\
         $G$ is \emph{not} $k$-connected & $O(k^3 n^2)$ & $O(k)$ & \ref{sec:cutvertex} \\
                  $G$ is (not) complete & {$O(n^2)$} & $O(1)$ & \ref{sec:examples:advanced} \\
         \midrule
         $G$ has a Hamilt. cycle (path)  & NP-complete & $O(1)$ & \ref{sec:hamilton} \\  
         $G$ has a length-$k$ cycle (path)  & NP-complete & $O(k)$ & \ref{sec:hamilton} \\
         \midrule
         $G$ is (not) bipartite &  {$O(n$+$m)$} & $O(1)$ & \ref{sec:examples:advanced} \\
         $G$ is $k$-colorable & NP-complete & $O(k)$ & \ref{sec:examples:advanced} \\
         \midrule
         CoNP-complete assertions & coNP-complete & \emph{Conj.:} No Visual Proof & \ref{sec:complexity}\\
         \bottomrule
    \end{tabular}
    \label{tab:overview}
\end{table}



\subparagraph{(Non)-Bipartiteness and $k$-colorability.}
\label{sec:colorability}

\begin{figure}[t]
    \centering
    \begin{subfigure}{0.24\linewidth}
    \centering
    \includegraphics[width=\linewidth]{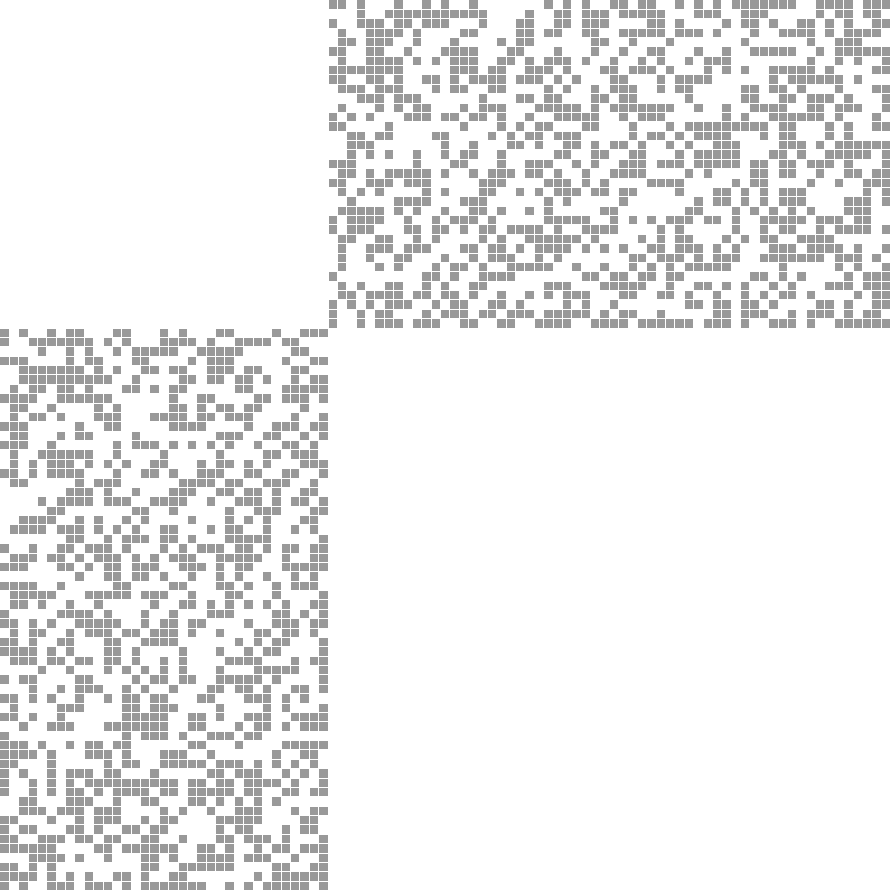}
    \subcaption{}
    \label{fig:bipartite}
    \end{subfigure}
    \hfill\begin{subfigure}{0.24\linewidth}
    \centering
    \includegraphics[width=\linewidth]{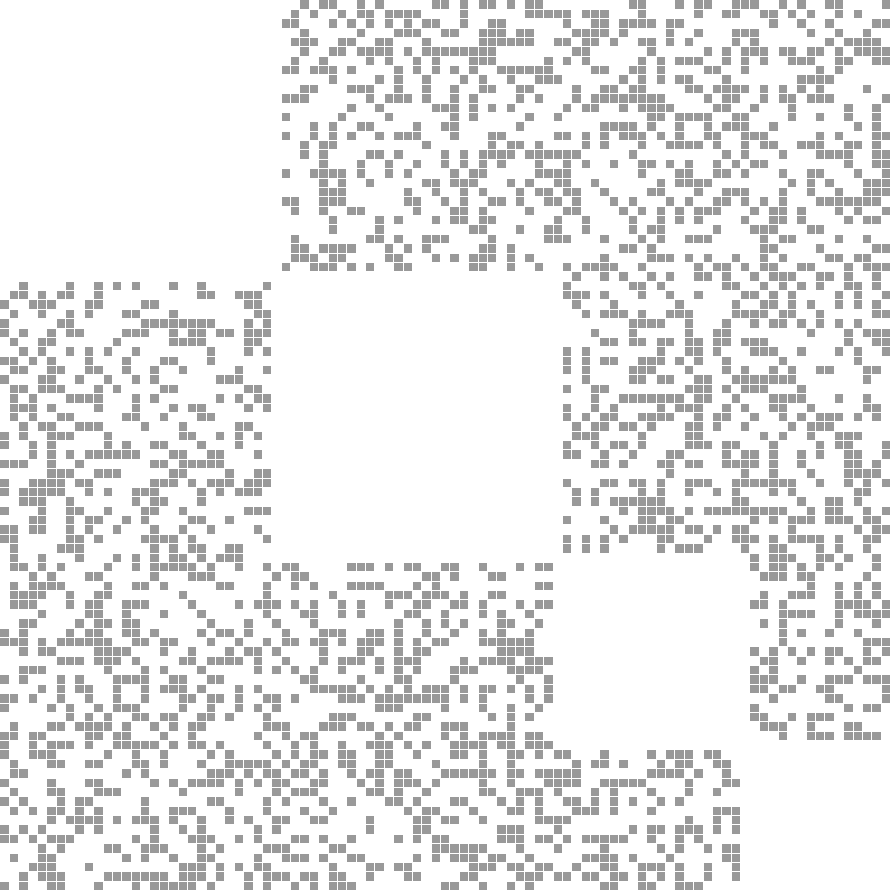}
    \subcaption{}
    \label{fig:coloring1}
    \end{subfigure}
    \hfill\begin{subfigure}{0.24\linewidth}
    \centering
    \includegraphics[width=\linewidth]{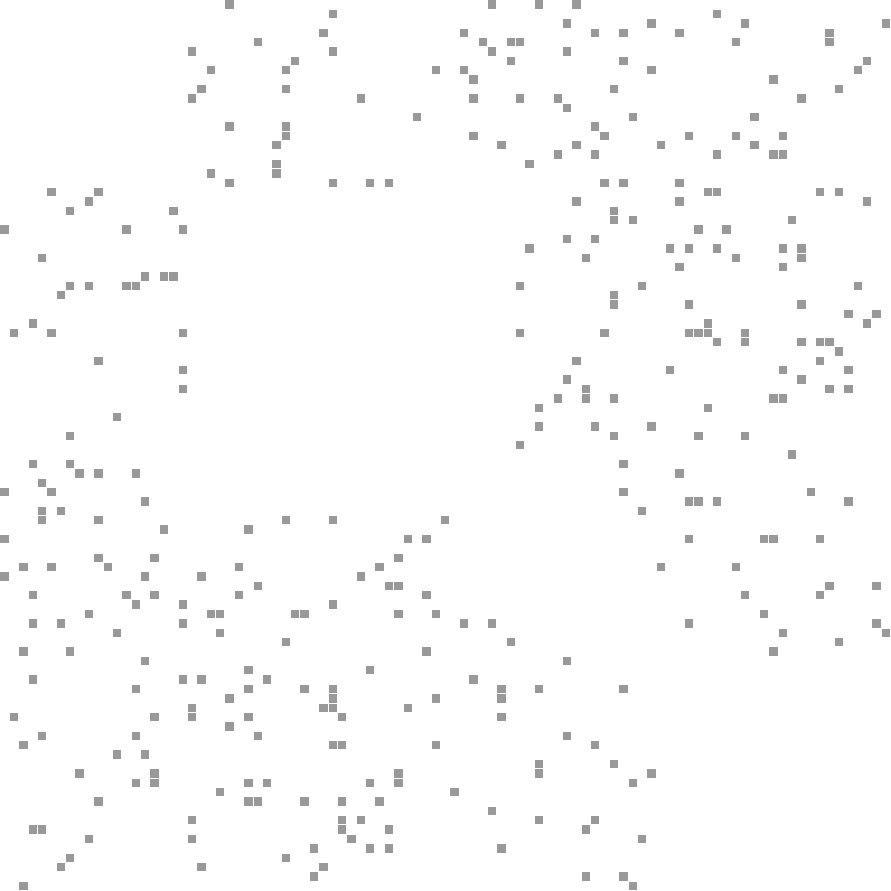}
    \subcaption{}
    \label{fig:coloring3}
    \end{subfigure}
    \hfill\begin{subfigure}{0.24\linewidth}
    \centering
    \includegraphics[width=\linewidth]{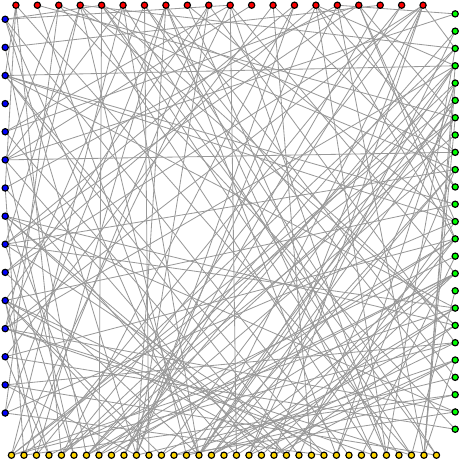}
    \subcaption{}
    \label{fig:coloring4}
    \end{subfigure}
    \caption{Visualizing $k$-colorability. (a)~A bipartite graph. (b)~A dense $4$-colorable graph.  (c) and (d) a adjacency matrix and a node-link visualization of a sparse $4$-colorable graph.}
    \label{fig:coloring}
\end{figure}

We can use a matrix representation to visually prove bipartiteness; see~\cref{fig:bipartite}.
When sorting the rows and columns according to the two independent subsets, bipartiteness can be simply checked by verifying if the two empty squares are indeed empty~\cite{Wolfe2010}. This approach also generalizes to $k$-colorability as shown in~\cref{fig:coloring1} for $4$ colors, however, for sparse graphs like in~\cref{fig:coloring3} additional highlighting of the (supposedly) empty squares might be necessary.
For small graphs, a node-link diagram 
 might be easier to read and hence preferable, however the approach does not scale well due to resolution since the judge needs to verify there are no edges within the subsets; see~\cref{fig:coloring4}.

\begin{figure}[t]
    \centering
    \begin{subfigure}{0.24\linewidth}
    \centering
    \includegraphics[width=\linewidth]{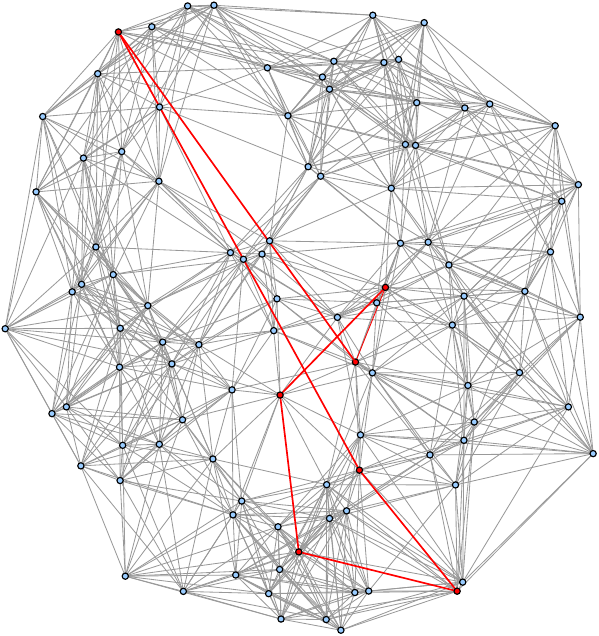}
    \subcaption{}
    \label{fig:nonbipartite0}
    \end{subfigure}
    \hfill\begin{subfigure}{0.24\linewidth}
    \centering
    \includegraphics[width=\linewidth]{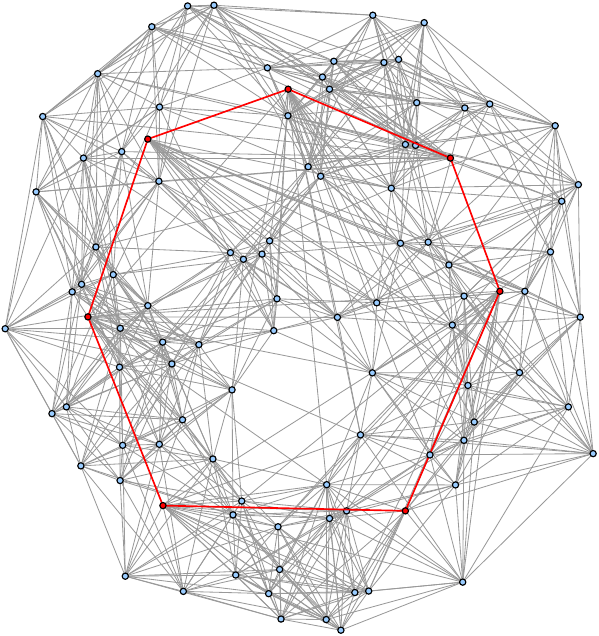}
    \subcaption{}
    \label{fig:nonbipartite2}
    \end{subfigure}
    \hfill\begin{subfigure}{0.24\linewidth}
    \centering
    \includegraphics[width=\linewidth]{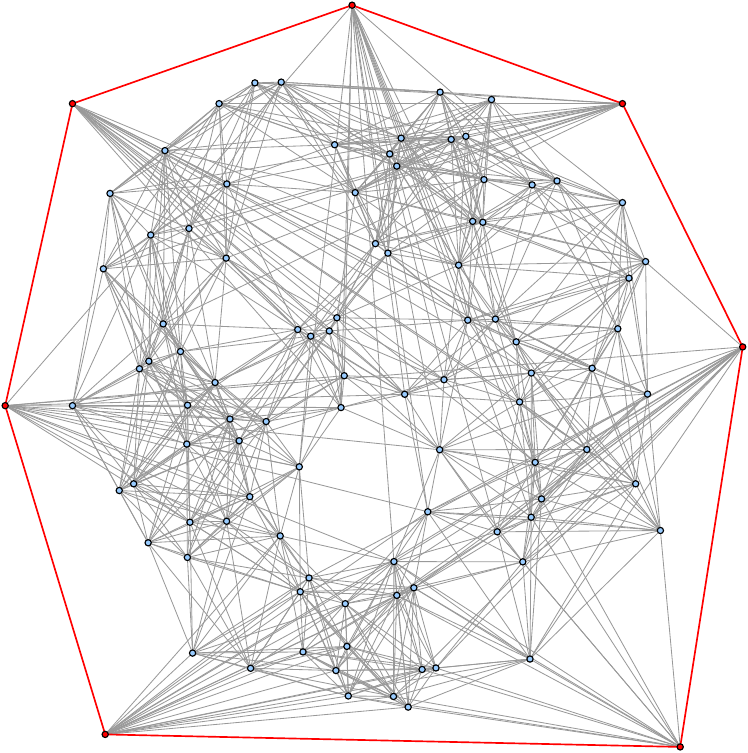}
    \subcaption{}
    \label{fig:nonbipartite3}
    \end{subfigure}
    \hfill\begin{subfigure}{0.24\linewidth}
    \centering
    \includegraphics[width=\linewidth]{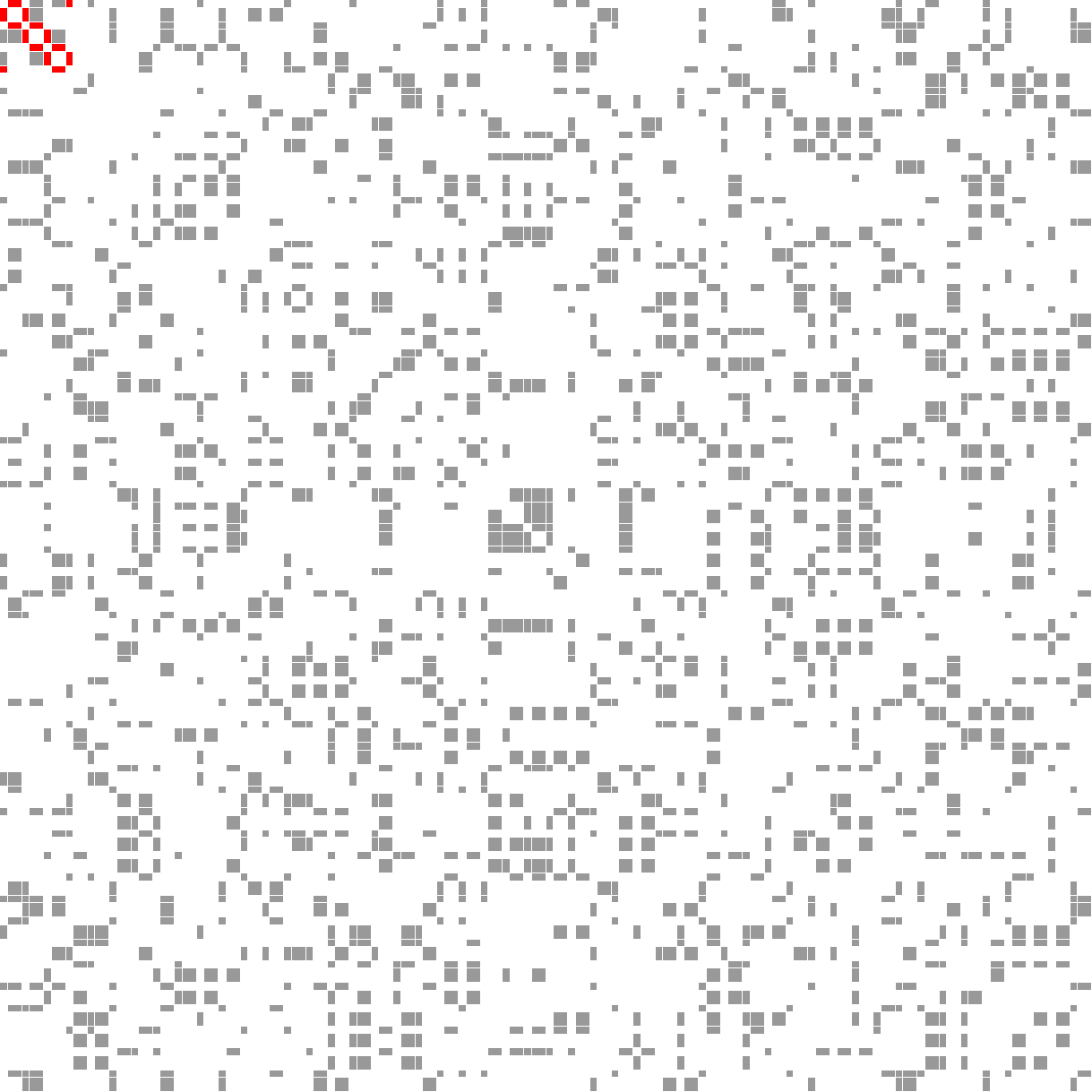}
    \subcaption{}
    \label{fig:nonbipartitematrix}
    \end{subfigure}
    \caption{Visualizing non-bipartiteness of a graph. In (a) the odd cycle is self-intersecting, making it difficult to certify that it is in fact a cycle. Both in (b) and (c) the cycle is clearly visibile where in (c) the cycle forms the outer boundary of the drawing letting it stand out even more compared to (b). Finally, in (d) the odd cycle is represented by a distinguishable pattern in the adjacency matrix.}
    \label{fig:nonbipartite}
\end{figure}

An odd-length cycle certifies that a graph is not bipartite, so non-bipartiteness can be visually proven by highlighting a shortest odd cycle in a drawing. In an arbitrary drawing, the cycle may be hard to spot, see~\cref{fig:nonbipartite0}.
Redrawing the cycle in convex position makes it easier to read (see~\cref{fig:nonbipartite2}), especially if it is the convex outer cycle; see~\cref{fig:nonbipartite3} (this  makes the rest of the graph harder to read; see \cref{tab:metrics}).
The cycle is now clearly visible and the judge just needs to assert oddness. While depending on the odd cycle length counting may be inevitable, the judge can use the symmetry of the drawing of the cycle to see that the cycle is odd (e.g., in \cref{fig:nonbipartite3}, there is a single top-most but no single bottom-most vertex). For larger cycle lengths, an adjacency matrix representation may be beneficial: 
Sort the rows and columns along the odd cycle and mark it, then append the remaining vertices arbitarily.
Then, alter the spacing of the matrix so that even rows and columns are thicker than odd ones; see \cref{fig:nonbipartitematrix}.
The cell closing the cycle is a square if and only if the length is odd.

\begin{figure}[t]
    \centering
    \begin{subfigure}{0.3\linewidth}
    \centering
    \includegraphics[width=\linewidth]{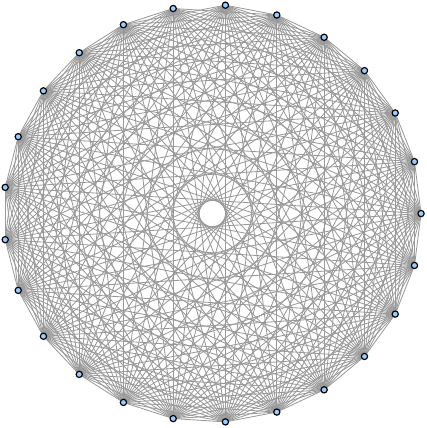}
    \subcaption{}
    \label{fig:noncomplete1}
    \end{subfigure}
    \hfill\begin{subfigure}{0.3\linewidth}
    \centering
    \includegraphics[width=\linewidth]{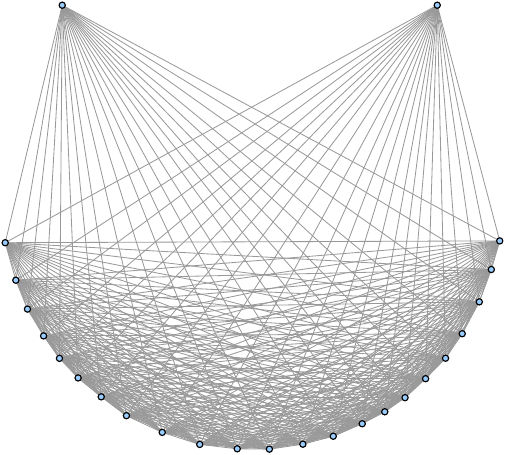}
    \subcaption{}
    \label{fig:noncomplete2}
    \end{subfigure}
    \hfill\begin{subfigure}{0.3\linewidth}
    \centering
    \includegraphics[width=\linewidth]{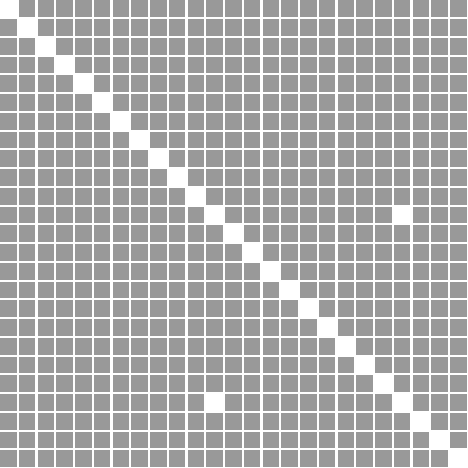}
    \subcaption{}
    \label{fig:noncomplete3}
    \end{subfigure}
    \caption{Visualizing non-completeness of a graph. In (a) the missing edge is very difficult to spot. In (b) and (c) on the other hand it is obvious that an edge is missing.}
    \label{fig:noncomplete}
\end{figure}

\subparagraph{Completeness and Non-Completeness.}
\label{sec:completeness}

Non-completeness is evidenced by a single missing edge and can be visually proven with a circular layout with the missing edge on the outer cycle. This approach does not scale well for a larger graphs; see~\cref{fig:noncomplete1}.
Readability and scalability can be improved by drawing focus to the missing edge, see~\cref{fig:noncomplete2}.
However, one can also use a matrix representation (see~\cref{fig:noncomplete3}) since spotting a missing square scales well from a perception perspective~\cite{Wolfe2010}. This technique can also prove completeness.

\section{Limitations of the \textsc{GraphTrials} Model}
\label{sec:limitations}


\subparagraph{Scalability.}

In the \textsc{GraphTrials} model, we must not only visualize the evidence represented in the visual certificate, but also display the remainder of the graph faithfully. This may result in higher computational complexity compared to other visualization techniques, e.\,g., force-directed graph layouts, whose purpose is to create an overall readable representation. 
Why not forgo visualization completely and use an assertion software to validate the evidence computationally? While this could drastically reduce the computation time and require fewer software components, there are in fact real-world application scenarios, e.\,g., in court, where it may be better to show a visual certificate accompanied by a short explanation why the certificate is indeed establishing the assertion instead of simply telling the audience that a piece of software analyzed the network and found the evidence for the assertion; see \cref{sec:intro}. Another benefit of visual proofs over a non-visual assertion software is that bugs in the visual proof pipeline can be spotted in the visual certificate, i.\,e., either the represented graph is not the input graph or the evidence is not a true evidence for the claim. 

Another scalability issue  is  to display the entire graph faithfully. In~\cref{sec:model}, we assumed that the visual certificate may be represented by few components in the judge's mental model and that the formation of that mental model can be mainly guided by usage of bottom-up and pattern recognition processes. For large input graphs, the screen resolution might not permit an information-faithful representation of the input graph so that one must resort to techniques for displaying larger data, e.\,g., zooming. The introduction of such modes of user interaction may be problematic for our model as it may lead the judge to increasingly use top-down processes of perception which may influence the formation of the mental model. 

\subparagraph{Human factors.} In our model, the judge is necessarily a human actor in the visual proof process. Hence, it is no surprise that human factors play an important role in the application of our model. Our model assumes that the judge is able to draw objective conclusions provided the evidence by the prosecution lawyer. This process may be hindered by insufficient background knowledge of the judge or subjective  expectations towards the visualization. Moreover, the judge's mental model cannot be directly analyzed and influenced introducing uncertainty into the model. We discuss these aspects further in~\cref{app:human}.

\section{Open Problems}
\label{sec:conclusion}


\begin{inlineenum}
    \item  Are visual proofs in fact scalable? How do they extend to geospatial and dynamic graphs where the data are expected to obey spatial and/or temporal constraints?
   \item Which features contribute to perceptual complexity? 
   \item Do response times depend mostly on perceptual complexity?
   \item When do human users regard a visual certificate as unimpeachable? 
   \item  What are human limits for the perception of graph properties? For instance, the minimum perceivable slope difference is $\approx 2$ degrees~\cite{tilt}.
   \item  What is the trade-off between perceptual complexity and cognitive load?
\end{inlineenum}

\bibliographystyle{abbrv-doi-hyperref}

\bibliography{references}

\begin{thebibliography}{10}

\bibitem{DBLP:journals/tvcg/AhmedLDKL22}
\href{https://doi.org/10.1109/TVCG.2022.3155564}{A.~R. Ahmed, F.~D. Luca,
  S.~Devkota, S.~G. Kobourov, and M.~Li}.
\newblock \href{https://doi.org/10.1109/TVCG.2022.3155564}{Multicriteria
  scalable graph drawing via stochastic gradient descent,
  {(SGD)\textsuperscript{2}}}.
\newblock \href{https://doi.org/10.1109/TVCG.2022.3155564}{{\em {IEEE} Trans.
  Vis. Comp. Graph.}},
  \href{https://doi.org/10.1109/TVCG.2022.3155564}{28(6):2388--2399},
  \href{https://doi.org/10.1109/TVCG.2022.3155564}{2022}.
  \href{https://doi.org/10.1109/TVCG.2022.3155564}
{doi: {{%
10\hspace{.1pt}\discretionary{.}{%
}{.}\hspace{.4pt}1109\discretionary{/}{%
}{/}TVCG\hspace{.1pt}\discretionary{.}{%
}{.}\hspace{.4pt}2022\hspace{.1pt}\discretionary{.}{%
}{.}\hspace{.4pt}3155564}}}


\bibitem{fiveColor}
\href{https://doi.org/https://doi.org/10.1007/978-3-662-05412-3_30}{M.~Aigner
  and G.~M. Ziegler}.
\newblock
  \href{https://doi.org/https://doi.org/10.1007/978-3-662-05412-3_30}{Five-coloring
  plane graphs}.
\newblock
  \href{https://doi.org/https://doi.org/10.1007/978-3-662-05412-3_30}{In {\em
  Proofs from {THE} {BOOK} {(3.} ed.)}}.
  \href{https://doi.org/https://doi.org/10.1007/978-3-662-05412-3_30}{Springer},
  \href{https://doi.org/https://doi.org/10.1007/978-3-662-05412-3_30}{2004}.
  \href{https://doi.org/10.1007/978-3-662-05412-3_30}
{doi: {{%
10\hspace{.1pt}\discretionary{.}{%
}{.}\hspace{.4pt}1007\discretionary{/}{%
}{/}978\discretionary{%
}{-}{-}3\discretionary{%
}{-}{-}662\discretionary{%
}{-}{-}05412\discretionary{%
}{-}{-}3\_30}}}


\bibitem{DBLP:conf/gd/AngoriDMPT19}
\href{https://doi.org/10.1007/978-3-030-35802-0_22}{L.~Angori, W.~Didimo,
  F.~Montecchiani, D.~Pagliuca, and A.~Tappini}.
\newblock \href{https://doi.org/10.1007/978-3-030-35802-0_22}{Chordlink: {A}
  new hybrid visualization model}.
\newblock \href{https://doi.org/10.1007/978-3-030-35802-0_22}{In D.~Archambault
  and C.~D. T{\'{o}}th, eds., {\em Graph Drawing and Network Visualization -
  27th International Symposium, {GD} 2019, Prague, Czech Republic, September
  17-20, 2019, Proceedings}},
  \href{https://doi.org/10.1007/978-3-030-35802-0_22}{vol. 11904 of {\em
  Lecture Notes in Computer Science}},
  \href{https://doi.org/10.1007/978-3-030-35802-0_22}{pp. 276--290}.
  \href{https://doi.org/10.1007/978-3-030-35802-0_22}{Springer},
  \href{https://doi.org/10.1007/978-3-030-35802-0_22}{2019}.
  \href{https://doi.org/10.1007/978-3-030-35802-0_22}
{doi: {{%
10\hspace{.1pt}\discretionary{.}{%
}{.}\hspace{.4pt}1007\discretionary{/}{%
}{/}978\discretionary{%
}{-}{-}3\discretionary{%
}{-}{-}030\discretionary{%
}{-}{-}35802\discretionary{%
}{-}{-}0\_22}}}


\bibitem{DBLP:journals/tvcg/AngoriDMPT22}
\href{https://doi.org/10.1109/TVCG.2020.3016055}{L.~Angori, W.~Didimo,
  F.~Montecchiani, D.~Pagliuca, and A.~Tappini}.
\newblock \href{https://doi.org/10.1109/TVCG.2020.3016055}{Hybrid graph
  visualizations with chordlink: Algorithms, experiments, and applications}.
\newblock \href{https://doi.org/10.1109/TVCG.2020.3016055}{{\em {IEEE} Trans.
  Vis. Comp. Graph.}},
  \href{https://doi.org/10.1109/TVCG.2020.3016055}{28(2):1288--1300},
  \href{https://doi.org/10.1109/TVCG.2020.3016055}{2022}.
  \href{https://doi.org/10.1109/TVCG.2020.3016055}
{doi: {{%
10\hspace{.1pt}\discretionary{.}{%
}{.}\hspace{.4pt}1109\discretionary{/}{%
}{/}TVCG\hspace{.1pt}\discretionary{.}{%
}{.}\hspace{.4pt}2020\hspace{.1pt}\discretionary{.}{%
}{.}\hspace{.4pt}3016055}}}


\bibitem{DBLP:journals/dm/AppelH76}
\href{https://doi.org/10.1016/0012-365X(76)90147-3}{K.~Appel and W.~Haken}.
\newblock \href{https://doi.org/10.1016/0012-365X(76)90147-3}{Special
  announcement}.
\newblock \href{https://doi.org/10.1016/0012-365X(76)90147-3}{{\em Discret.
  Math.}}, \href{https://doi.org/10.1016/0012-365X(76)90147-3}{16(2):179--180},
  \href{https://doi.org/10.1016/0012-365X(76)90147-3}{1976}.
  \href{https://doi.org/10.1016/0012-365X(76)90147-3}
{doi: {{%
10\hspace{.1pt}\discretionary{.}{%
}{.}\hspace{.4pt}1016\discretionary{/}{%
}{/}0012\discretionary{%
}{-}{-}365X\discretionary{%
}{(}{(}76\discretionary{)}{%
}{)}90147\discretionary{%
}{-}{-}3}}}


\bibitem{conp}
\href{https://doi.org/https://doi.org/10.1017/CBO9780511804090}{S.~Arora and
  B.~Barak}.
\newblock \href{https://doi.org/https://doi.org/10.1017/CBO9780511804090}{{\em
  Computational Complexity: A Modern Approach}}.
\newblock
  \href{https://doi.org/https://doi.org/10.1017/CBO9780511804090}{Cambridge
  University Press},
  \href{https://doi.org/https://doi.org/10.1017/CBO9780511804090}{2009}.
  \href{https://doi.org/10.1017/CBO9780511804090}
{doi: {{%
10\hspace{.1pt}\discretionary{.}{%
}{.}\hspace{.4pt}1017\discretionary{/}{%
}{/}CBO9780511804090}}}


\bibitem{DBLP:journals/cgf/BehrischBRSF16}
\href{https://doi.org/10.1111/cgf.12935}{M.~Behrisch, B.~Bach, N.~H. Riche,
  T.~Schreck, and J.~Fekete}.
\newblock \href{https://doi.org/10.1111/cgf.12935}{Matrix reordering methods
  for table and network visualization}.
\newblock \href{https://doi.org/10.1111/cgf.12935}{{\em Comput. Graph. Forum}},
  \href{https://doi.org/10.1111/cgf.12935}{35(3):693--716},
  \href{https://doi.org/10.1111/cgf.12935}{2016}.
  \href{https://doi.org/10.1111/cgf.12935}
{doi: {{%
10\hspace{.1pt}\discretionary{.}{%
}{.}\hspace{.4pt}1111\discretionary{/}{%
}{/}cgf\hspace{.1pt}\discretionary{.}{%
}{.}\hspace{.4pt}12935}}}


\bibitem{DBLP:conf/gd/BekosFGH0SS18}
\href{https://doi.org/10.1007/978-3-030-04414-5_19}{M.~A. Bekos,
  H.~F{\"{o}}rster, C.~Geckeler, L.~Holl{\"{a}}nder, M.~Kaufmann, A.~M.
  Spallek, and J.~Splett}.
\newblock \href{https://doi.org/10.1007/978-3-030-04414-5_19}{A heuristic
  approach towards drawings of graphs with high crossing resolution}.
\newblock \href{https://doi.org/10.1007/978-3-030-04414-5_19}{In T.~Biedl and
  A.~Kerren, eds., {\em Graph Drawing and Network Visualization - 26th
  International Symposium, {GD} 2018, Barcelona, Spain, September 26-28, 2018,
  Proceedings}}, \href{https://doi.org/10.1007/978-3-030-04414-5_19}{vol. 11282
  of {\em Lecture Notes in Computer Science}},
  \href{https://doi.org/10.1007/978-3-030-04414-5_19}{pp. 271--285}.
  \href{https://doi.org/10.1007/978-3-030-04414-5_19}{Springer},
  \href{https://doi.org/10.1007/978-3-030-04414-5_19}{2018}.
  \href{https://doi.org/10.1007/978-3-030-04414-5_19}
{doi: {{%
10\hspace{.1pt}\discretionary{.}{%
}{.}\hspace{.4pt}1007\discretionary{/}{%
}{/}978\discretionary{%
}{-}{-}3\discretionary{%
}{-}{-}030\discretionary{%
}{-}{-}04414\discretionary{%
}{-}{-}5\_19}}}


\bibitem{DBLP:journals/cj/BekosFGHKSS21}
\href{https://doi.org/10.1093/comjnl/bxz133}{M.~A. Bekos, H.~F{\"{o}}rster,
  C.~Geckeler, L.~Holl{\"{a}}nder, M.~Kaufmann, A.~M. Spallek, and J.~Splett}.
\newblock \href{https://doi.org/10.1093/comjnl/bxz133}{A heuristic approach
  towards drawings of graphs with high crossing resolution}.
\newblock \href{https://doi.org/10.1093/comjnl/bxz133}{{\em Comput. J.}},
  \href{https://doi.org/10.1093/comjnl/bxz133}{64(1):7--26},
  \href{https://doi.org/10.1093/comjnl/bxz133}{2021}.
  \href{https://doi.org/10.1093/comjnl/bxz133}
{doi: {{%
10\hspace{.1pt}\discretionary{.}{%
}{.}\hspace{.4pt}1093\discretionary{/}{%
}{/}comjnl\discretionary{/}{%
}{/}bxz133}}}


\bibitem{DBLP:conf/iclr/BelloPL0B17}
\href{https://openreview.net/forum?id=Bk9mxlSFx}{I.~Bello, H.~Pham, Q.~V. Le,
  M.~Norouzi, and S.~Bengio}.
\newblock \href{https://openreview.net/forum?id=Bk9mxlSFx}{Neural combinatorial
  optimization with reinforcement learning}.
\newblock \href{https://openreview.net/forum?id=Bk9mxlSFx}{In {\em
  Intl.~Conf.~Learning Represent. {ICLR}}}.
  \href{https://openreview.net/forum?id=Bk9mxlSFx}{OpenReview.net},
  \href{https://openreview.net/forum?id=Bk9mxlSFx}{2017}.

\bibitem{narratingnetworks}
\href{https://doi.org/https://doi.org/10.1080/21670811.2016.1186497}{L.~Bounegru,
  T.~Venturini, J.~Gray, and M.~Jacomy}.
\newblock
  \href{https://doi.org/https://doi.org/10.1080/21670811.2016.1186497}{Narrating
  networks}.
\newblock
  \href{https://doi.org/https://doi.org/10.1080/21670811.2016.1186497}{{\em
  Digital Journalism}},
  \href{https://doi.org/https://doi.org/10.1080/21670811.2016.1186497}{5(6):699--730},
  \href{https://doi.org/https://doi.org/10.1080/21670811.2016.1186497}{2017}.
  \href{https://doi.org/10.1080/21670811.2016.1186497}
{doi: {{%
10\hspace{.1pt}\discretionary{.}{%
}{.}\hspace{.4pt}1080\discretionary{/}{%
}{/}21670811\hspace{.1pt}\discretionary{.}{%
}{.}\hspace{.4pt}2016\hspace{.1pt}\discretionary{.}{%
}{.}\hspace{.4pt}1186497}}}


\bibitem{philosophy}
\href{https://doi.org/https://doi.org/10.4324/9780203932964}{J.~R. Brown}.
\newblock \href{https://doi.org/https://doi.org/10.4324/9780203932964}{{\em
  Philosophy of Mathematics: A Contemporary Introduction to the World of Proofs
  and Pictures}}.
\newblock
  \href{https://doi.org/https://doi.org/10.4324/9780203932964}{Routledge},
  \href{https://doi.org/https://doi.org/10.4324/9780203932964}{2nd ed.},
  \href{https://doi.org/https://doi.org/10.4324/9780203932964}{2008}.
  \href{https://doi.org/10.4324/9780203932964}
{doi: {{%
10\hspace{.1pt}\discretionary{.}{%
}{.}\hspace{.4pt}4324\discretionary{/}{%
}{/}9780203932964}}}


\bibitem{DBLP:conf/vinci/BurchBCKPSV20}
\href{https://doi.org/10.1145/3430036.3430064}{M.~Burch, K.~B. ten Brinke,
  A.~Castella, G.~Karray, S.~Peters, V.~Shteriyanov, and R.~Vlasvinkel}.
\newblock \href{https://doi.org/10.1145/3430036.3430064}{Guiding graph
  exploration by combining layouts and reorderings}.
\newblock \href{https://doi.org/10.1145/3430036.3430064}{In M.~Burch, M.~A.
  Westenberg, Q.~V. Nguyen, and Y.~Zhao, eds., {\em {VINCI}: Intl.~Symp.~Vis.
  Inform. Comm. Interact.}}, \href{https://doi.org/10.1145/3430036.3430064}{pp.
  25:1--25:5}. \href{https://doi.org/10.1145/3430036.3430064}{{ACM}},
  \href{https://doi.org/10.1145/3430036.3430064}{2020}.
  \href{https://doi.org/10.1145/3430036.3430064}
{doi: {{%
10\hspace{.1pt}\discretionary{.}{%
}{.}\hspace{.4pt}1145\discretionary{/}{%
}{/}3430036\hspace{.1pt}\discretionary{.}{%
}{.}\hspace{.4pt}3430064}}}


\bibitem{card1999readings}
\href{https://doi.org/https://doi.org/10.1111/cgf.13730}{S.~K. Card, J.~D.
  Mackinlay, and B.~Shneiderman}.
\newblock \href{https://doi.org/https://doi.org/10.1111/cgf.13730}{{\em
  Readings in information visualization: using vision to think}}.
\newblock \href{https://doi.org/https://doi.org/10.1111/cgf.13730}{Morgan
  Kaufmann Publishers Inc.},
  \href{https://doi.org/https://doi.org/10.1111/cgf.13730}{San Francisco, CA,
  USA}, \href{https://doi.org/https://doi.org/10.1111/cgf.13730}{1999}.
  \href{https://doi.org/10.1111/cgf.13730}
{doi: {{%
10\hspace{.1pt}\discretionary{.}{%
}{.}\hspace{.4pt}1111\discretionary{/}{%
}{/}cgf\hspace{.1pt}\discretionary{.}{%
}{.}\hspace{.4pt}13730}}}


\bibitem{DBLP:journals/cacm/CardMN80}
\href{https://doi.org/10.1145/358886.358895}{S.~K. Card, T.~P. Moran, and
  A.~Newell}.
\newblock \href{https://doi.org/10.1145/358886.358895}{The keystroke-level
  model for user performance time with interactice systems}.
\newblock \href{https://doi.org/10.1145/358886.358895}{{\em Commun. {ACM}}},
  \href{https://doi.org/10.1145/358886.358895}{23(7):396--410},
  \href{https://doi.org/10.1145/358886.358895}{1980}.
  \href{https://doi.org/10.1145/358886.358895}
{doi: {{%
10\hspace{.1pt}\discretionary{.}{%
}{.}\hspace{.4pt}1145\discretionary{/}{%
}{/}358886\hspace{.1pt}\discretionary{.}{%
}{.}\hspace{.4pt}358895}}}


\bibitem{DBLP:books/lib/CardMN83}
\href{https://www.worldcat.org/oclc/59953542}{S.~K. Card, T.~P. Moran, and
  A.~Newell}.
\newblock \href{https://www.worldcat.org/oclc/59953542}{{\em The psychology of
  human-computer interaction}}.
\newblock \href{https://www.worldcat.org/oclc/59953542}{Erlbaum},
  \href{https://www.worldcat.org/oclc/59953542}{1983}.

\bibitem{ceneda2019review}
D.~Ceneda, T.~Gschwandtner, and S.~Miksch.
\newblock A review of guidance approaches in visual data analysis: A multifocal
  perspective.
\newblock {\em Comput. Graph. Forum}, 38(3):861--879, 2019.

\bibitem{AIHPB_1967__3_4_433_0}
\href{http://www.numdam.org/item/AIHPB_1967__3_4_433_0/}{G.~Chartrand and
  F.~Harary}.
\newblock \href{http://www.numdam.org/item/AIHPB_1967__3_4_433_0/}{Planar
  {Permutation} {Graphs}}.
\newblock \href{http://www.numdam.org/item/AIHPB_1967__3_4_433_0/}{{\em Annales
  de l'institut Henri Poincar\'e. Section B. Calcul des probabilit\'es et
  statistiques}},
  \href{http://www.numdam.org/item/AIHPB_1967__3_4_433_0/}{3(4):433--438},
  \href{http://www.numdam.org/item/AIHPB_1967__3_4_433_0/}{1967}.

\bibitem{DBLP:conf/ipco/CheriyanSS98}
\href{https://doi.org/10.1007/3-540-69346-7_10}{J.~Cheriyan, A.~Seb{\"{o}}, and
  Z.~Szigeti}.
\newblock \href{https://doi.org/10.1007/3-540-69346-7_10}{An improved
  approximation algorithm for minimum size 2-edge connected spanning
  subgraphs}.
\newblock \href{https://doi.org/10.1007/3-540-69346-7_10}{In R.~E. Bixby, E.~A.
  Boyd, and R.~Z. R{\'{\i}}os{-}Mercado, eds., {\em Integer Programming and
  Combinatorial Optimization, Intl. {IPCO} Conf.~Proc.}},
  \href{https://doi.org/10.1007/3-540-69346-7_10}{vol. 1412 of {\em LNCS}},
  \href{https://doi.org/10.1007/3-540-69346-7_10}{pp. 126--136}.
  \href{https://doi.org/10.1007/3-540-69346-7_10}{Springer},
  \href{https://doi.org/10.1007/3-540-69346-7_10}{1998}.
  \href{https://doi.org/10.1007/3-540-69346-7_10}
{doi: {{%
10\hspace{.1pt}\discretionary{.}{%
}{.}\hspace{.4pt}1007\discretionary{/}{%
}{/}3\discretionary{%
}{-}{-}540\discretionary{%
}{-}{-}69346\discretionary{%
}{-}{-}7\_10}}}


\bibitem{DBLP:journals/cga/ChooL18}
\href{https://doi.org/10.1109/MCG.2018.042731661}{J.~Choo and S.~Liu}.
\newblock \href{https://doi.org/10.1109/MCG.2018.042731661}{Visual analytics
  for explainable deep learning}.
\newblock \href{https://doi.org/10.1109/MCG.2018.042731661}{{\em {IEEE} Comput.
  Graph. Appl.}},
  \href{https://doi.org/10.1109/MCG.2018.042731661}{38(4):84--92},
  \href{https://doi.org/10.1109/MCG.2018.042731661}{2018}.
  \href{https://doi.org/10.1109/MCG.2018.042731661}
{doi: {{%
10\hspace{.1pt}\discretionary{.}{%
}{.}\hspace{.4pt}1109\discretionary{/}{%
}{/}MCG\hspace{.1pt}\discretionary{.}{%
}{.}\hspace{.4pt}2018\hspace{.1pt}\discretionary{.}{%
}{.}\hspace{.4pt}042731661}}}


\bibitem{DBLP:conf/gd/DevkotaALIK19}
\href{https://doi.org/10.1007/978-3-030-35802-0_23}{S.~Devkota, A.~R. Ahmed,
  F.~D. Luca, K.~E. Isaacs, and S.~G. Kobourov}.
\newblock \href{https://doi.org/10.1007/978-3-030-35802-0_23}{Stress-plus-x
  {(SPX)} graph layout}.
\newblock \href{https://doi.org/10.1007/978-3-030-35802-0_23}{In D.~Archambault
  and C.~D. T{\'{o}}th, eds., {\em Intl.~Symp.~Graph Drawing}},
  \href{https://doi.org/10.1007/978-3-030-35802-0_23}{vol. 11904 of {\em
  LNCS}}, \href{https://doi.org/10.1007/978-3-030-35802-0_23}{pp. 291--304}.
  \href{https://doi.org/10.1007/978-3-030-35802-0_23}{Springer},
  \href{https://doi.org/10.1007/978-3-030-35802-0_23}{2019}.
  \href{https://doi.org/10.1007/978-3-030-35802-0_23}
{doi: {{%
10\hspace{.1pt}\discretionary{.}{%
}{.}\hspace{.4pt}1007\discretionary{/}{%
}{/}978\discretionary{%
}{-}{-}3\discretionary{%
}{-}{-}030\discretionary{%
}{-}{-}35802\discretionary{%
}{-}{-}0\_23}}}


\bibitem{DBLP:books/ph/BattistaETT99}
G.~{Di Battista}, P.~Eades, R.~Tamassia, and I.~G. Tollis.
\newblock {\em Graph Drawing: Algorithms for the Visualization of Graphs}.
\newblock Prentice-Hall, 1999.

\bibitem{DBLP:conf/gd/GiacomoDLM09}
\href{https://doi.org/10.1007/978-3-642-11805-0_4}{E.~{Di Giacomo}, W.~Didimo,
  G.~Liotta, and H.~Meijer}.
\newblock \href{https://doi.org/10.1007/978-3-642-11805-0_4}{Area, curve
  complexity, and crossing resolution of non-planar graph drawings}.
\newblock \href{https://doi.org/10.1007/978-3-642-11805-0_4}{In D.~Eppstein and
  E.~R. Gansner, eds., {\em Graph Drawing, 17th International Symposium, {GD}
  2009, Chicago, IL, USA, September 22-25, 2009. Revised Papers}},
  \href{https://doi.org/10.1007/978-3-642-11805-0_4}{vol. 5849 of {\em Lecture
  Notes in Computer Science}},
  \href{https://doi.org/10.1007/978-3-642-11805-0_4}{pp. 15--20}.
  \href{https://doi.org/10.1007/978-3-642-11805-0_4}{Springer},
  \href{https://doi.org/10.1007/978-3-642-11805-0_4}{2009}.
  \href{https://doi.org/10.1007/978-3-642-11805-0_4}
{doi: {{%
10\hspace{.1pt}\discretionary{.}{%
}{.}\hspace{.4pt}1007\discretionary{/}{%
}{/}978\discretionary{%
}{-}{-}3\discretionary{%
}{-}{-}642\discretionary{%
}{-}{-}11805\discretionary{%
}{-}{-}0\_4}}}


\bibitem{DBLP:journals/mst/GiacomoDLM11}
\href{https://doi.org/10.1007/s00224-010-9275-6}{E.~{Di Giacomo}, W.~Didimo,
  G.~Liotta, and H.~Meijer}.
\newblock \href{https://doi.org/10.1007/s00224-010-9275-6}{Area, curve
  complexity, and crossing resolution of non-planar graph drawings}.
\newblock \href{https://doi.org/10.1007/s00224-010-9275-6}{{\em Theory Comput.
  Syst.}}, \href{https://doi.org/10.1007/s00224-010-9275-6}{49(3):565--575},
  \href{https://doi.org/10.1007/s00224-010-9275-6}{2011}.
  \href{https://doi.org/10.1007/s00224-010-9275-6}
{doi: {{%
10\hspace{.1pt}\discretionary{.}{%
}{.}\hspace{.4pt}1007\discretionary{/}{%
}{/}s00224\discretionary{%
}{-}{-}010\discretionary{%
}{-}{-}9275\discretionary{%
}{-}{-}6}}}


\bibitem{DBLP:journals/dss/DidimoGLMP18}
\href{https://doi.org/10.1016/j.dss.2018.03.008}{W.~Didimo, L.~Giamminonni,
  G.~Liotta, F.~Montecchiani, and D.~Pagliuca}.
\newblock \href{https://doi.org/10.1016/j.dss.2018.03.008}{A visual analytics
  system to support tax evasion discovery}.
\newblock \href{https://doi.org/10.1016/j.dss.2018.03.008}{{\em Decis. Support
  Syst.}}, \href{https://doi.org/10.1016/j.dss.2018.03.008}{110:71--83},
  \href{https://doi.org/10.1016/j.dss.2018.03.008}{2018}.
  \href{https://doi.org/10.1016/j.dss.2018.03.008}
{doi: {{%
10\hspace{.1pt}\discretionary{.}{%
}{.}\hspace{.4pt}1016\discretionary{/}{%
}{/}j\hspace{.1pt}\discretionary{.}{%
}{.}\hspace{.4pt}dss\hspace{.1pt}\discretionary{.}{%
}{.}\hspace{.4pt}2018\hspace{.1pt}\discretionary{.}{%
}{.}\hspace{.4pt}03\hspace{.1pt}\discretionary{.}{%
}{.}\hspace{.4pt}008}}}


\bibitem{DBLP:journals/access/DidimoGLMMP20}
\href{https://doi.org/10.1109/ACCESS.2020.2967974}{W.~Didimo, L.~Grilli,
  G.~Liotta, L.~Menconi, F.~Montecchiani, and D.~Pagliuca}.
\newblock \href{https://doi.org/10.1109/ACCESS.2020.2967974}{Combining network
  visualization and data mining for tax risk assessment}.
\newblock \href{https://doi.org/10.1109/ACCESS.2020.2967974}{{\em {IEEE}
  Access}}, \href{https://doi.org/10.1109/ACCESS.2020.2967974}{8:16073--16086},
  \href{https://doi.org/10.1109/ACCESS.2020.2967974}{2020}.
  \href{https://doi.org/10.1109/ACCESS.2020.2967974}
{doi: {{%
10\hspace{.1pt}\discretionary{.}{%
}{.}\hspace{.4pt}1109\discretionary{/}{%
}{/}ACCESS\hspace{.1pt}\discretionary{.}{%
}{.}\hspace{.4pt}2020\hspace{.1pt}\discretionary{.}{%
}{.}\hspace{.4pt}2967974}}}


\bibitem{DBLP:journals/isci/DidimoGLMP19}
\href{https://doi.org/10.1016/j.ins.2019.07.097}{W.~Didimo, L.~Grilli,
  G.~Liotta, F.~Montecchiani, and D.~Pagliuca}.
\newblock
  \href{https://doi.org/10.1016/j.ins.2019.07.097}{Visual~querying~and~ana\-lysis
  of temporal fiscal networks}.
\newblock \href{https://doi.org/10.1016/j.ins.2019.07.097}{{\em Inf.~Sci.}},
  \href{https://doi.org/10.1016/j.ins.2019.07.097}{505:406--421},
  \href{https://doi.org/10.1016/j.ins.2019.07.097}{2019}.
  \href{https://doi.org/10.1016/j.ins.2019.07.097}
{doi: {{%
10\hspace{.1pt}\discretionary{.}{%
}{.}\hspace{.4pt}1016\discretionary{/}{%
}{/}j\hspace{.1pt}\discretionary{.}{%
}{.}\hspace{.4pt}ins\hspace{.1pt}\discretionary{.}{%
}{.}\hspace{.4pt}2019\hspace{.1pt}\discretionary{.}{%
}{.}\hspace{.4pt}07\hspace{.1pt}\discretionary{.}{%
}{.}\hspace{.4pt}097}}}


\bibitem{DBLP:books/fm/GareyJ79}
M.~R. Garey and D.~S. Johnson.
\newblock {\em Computers and Intractability: A Guide to the Theory of
  NP-Completeness}.
\newblock W. H. Freeman, 1979.

\bibitem{DBLP:journals/tcs/GeorgiadisIK20}
\href{https://doi.org/10.1016/j.tcs.2019.09.040}{L.~Georgiadis, G.~F. Italiano,
  and A.~Karanasiou}.
\newblock \href{https://doi.org/10.1016/j.tcs.2019.09.040}{Approximating the
  smallest 2-vertex connected spanning subgraph of a directed graph}.
\newblock \href{https://doi.org/10.1016/j.tcs.2019.09.040}{{\em Theoret.
  Comput. Science}},
  \href{https://doi.org/10.1016/j.tcs.2019.09.040}{807:185--200},
  \href{https://doi.org/10.1016/j.tcs.2019.09.040}{2020}.
  \href{https://doi.org/10.1016/j.tcs.2019.09.040}
{doi: {{%
10\hspace{.1pt}\discretionary{.}{%
}{.}\hspace{.4pt}1016\discretionary{/}{%
}{/}j\hspace{.1pt}\discretionary{.}{%
}{.}\hspace{.4pt}tcs\hspace{.1pt}\discretionary{.}{%
}{.}\hspace{.4pt}2019\hspace{.1pt}\discretionary{.}{%
}{.}\hspace{.4pt}09\hspace{.1pt}\discretionary{.}{%
}{.}\hspace{.4pt}040}}}


\bibitem{GOLDSCHMIDT199697}
\href{https://doi.org/https://doi.org/10.1016/0377-2217(95)00202-2}{O.~Goldschmidt,
  A.~Takvorian, and G.~Yu}.
\newblock
  \href{https://doi.org/https://doi.org/10.1016/0377-2217(95)00202-2}{On
  finding a biconnected spanning planar subgraph with applications to the
  facilities layout problem}.
\newblock
  \href{https://doi.org/https://doi.org/10.1016/0377-2217(95)00202-2}{{\em
  Europ. J. Operat. Res.}},
  \href{https://doi.org/https://doi.org/10.1016/0377-2217(95)00202-2}{94(1):97--105},
  \href{https://doi.org/https://doi.org/10.1016/0377-2217(95)00202-2}{1996}.
  \href{https://doi.org/10.1016/0377-2217(95)00202-2}
{doi: {{%
10\hspace{.1pt}\discretionary{.}{%
}{.}\hspace{.4pt}1016\discretionary{/}{%
}{/}0377\discretionary{%
}{-}{-}2217\discretionary{%
}{(}{(}95\discretionary{)}{%
}{)}00202\discretionary{%
}{-}{-}2}}}


\bibitem{DBLP:conf/iv/Gove19}
\href{https://doi.org/10.1109/IV.2019.00042}{R.~Gove}.
\newblock \href{https://doi.org/10.1109/IV.2019.00042}{Gragnostics: Fast,
  interpretable features for comparing graphs}.
\newblock \href{https://doi.org/10.1109/IV.2019.00042}{In E.~Banissi, A.~Ursyn,
  M.~W.~M. Bannatyne, N.~Datia, R.~Francese, M.~Sarfraz, T.~G. Wyeld,
  F.~Bouali, G.~Venturini, H.~Azzag, M.~Lebbah, M.~Trutschl, U.~Cvek,
  H.~M{\"{u}}ller, M.~Nakayama, S.~Kernbach, L.~Caruccio, M.~Risi, U.~Erra,
  A.~Vitiello, and V.~Rossano, eds., {\em Intl. Conf. Inform. Visual. {IV}}},
  \href{https://doi.org/10.1109/IV.2019.00042}{pp. 201--209}.
  \href{https://doi.org/10.1109/IV.2019.00042}{{IEEE}},
  \href{https://doi.org/10.1109/IV.2019.00042}{2019}.
  \href{https://doi.org/10.1109/IV.2019.00042}
{doi: {{%
10\hspace{.1pt}\discretionary{.}{%
}{.}\hspace{.4pt}1109\discretionary{/}{%
}{/}IV\hspace{.1pt}\discretionary{.}{%
}{.}\hspace{.4pt}2019\hspace{.1pt}\discretionary{.}{%
}{.}\hspace{.4pt}00042}}}


\bibitem{Gove2022}
\href{https://doi.org/10.1007/978-3-030-93119-3_12}{R.~Gove}.
\newblock \href{https://doi.org/10.1007/978-3-030-93119-3_12}{Gragnostics:
  Evaluating fast, interpretable structural graph features for classification
  and visual analytics}.
\newblock \href{https://doi.org/10.1007/978-3-030-93119-3_12}{In
  B.~Kovalerchuk, K.~Nazemi, R.~Andonie, N.~Datia, and E.~Banissi, eds., {\em
  Integrating Artificial Intelligence and Visualization for Visual Knowledge
  Discovery}}, \href{https://doi.org/10.1007/978-3-030-93119-3_12}{pp.
  311--336}. \href{https://doi.org/10.1007/978-3-030-93119-3_12}{Springer},
  \href{https://doi.org/10.1007/978-3-030-93119-3_12}{2022}.
  \href{https://doi.org/10.1007/978-3-030-93119-3_12}
{doi: {{%
10\hspace{.1pt}\discretionary{.}{%
}{.}\hspace{.4pt}1007\discretionary{/}{%
}{/}978\discretionary{%
}{-}{-}3\discretionary{%
}{-}{-}030\discretionary{%
}{-}{-}93119\discretionary{%
}{-}{-}3\_12}}}


\bibitem{Grzeczkowski2017}
\href{https://doi.org/10.1016/j.visres.2016.10.006}{L.~Grzeczkowski, A.~M.
  Clarke, G.~Francis, F.~W. Mast, and M.~H. Herzog}.
\newblock \href{https://doi.org/10.1016/j.visres.2016.10.006}{{About individual
  differences in vision}}.
\newblock \href{https://doi.org/10.1016/j.visres.2016.10.006}{{\em Vision
  Research}},
  \href{https://doi.org/10.1016/j.visres.2016.10.006}{141:282--292},
  \href{https://doi.org/10.1016/j.visres.2016.10.006}{2017}.
  \href{https://doi.org/10.1016/j.visres.2016.10.006}
{doi: {{%
10\hspace{.1pt}\discretionary{.}{%
}{.}\hspace{.4pt}1016\discretionary{/}{%
}{/}j\hspace{.1pt}\discretionary{.}{%
}{.}\hspace{.4pt}visres\hspace{.1pt}\discretionary{.}{%
}{.}\hspace{.4pt}2016\hspace{.1pt}\discretionary{.}{%
}{.}\hspace{.4pt}10\hspace{.1pt}\discretionary{.}{%
}{.}\hspace{.4pt}006}}}


\bibitem{Healy2018}
K.~Healy.
\newblock {\em {Data Visualization: A Practical Introduction}}.
\newblock Princeton University Press, 2018.

\bibitem{HR92}
\href{https://doi.org/10.1137/0221055}{L.~S. Heath and A.~L. Rosenberg}.
\newblock \href{https://doi.org/10.1137/0221055}{Laying out graphs using
  queues}.
\newblock \href{https://doi.org/10.1137/0221055}{{\em {SIAM} J. Comput.}},
  \href{https://doi.org/10.1137/0221055}{21(5):927--958},
  \href{https://doi.org/10.1137/0221055}{1992}.
  \href{https://doi.org/10.1137/0221055}
{doi: {{%
10\hspace{.1pt}\discretionary{.}{%
}{.}\hspace{.4pt}1137\discretionary{/}{%
}{/}0221055}}}


\bibitem{DBLP:journals/tvcg/HenryF06}
\href{https://doi.org/10.1109/TVCG.2006.160}{N.~Henry and J.~Fekete}.
\newblock
  \href{https://doi.org/10.1109/TVCG.2006.160}{{MatrixExplorer}:~a~dual-representation~system~to~explore~social~net\-works}.
\newblock \href{https://doi.org/10.1109/TVCG.2006.160}{{\em IEEE
  Trans.~Vis.~Comp.~Graph.}},
  \href{https://doi.org/10.1109/TVCG.2006.160}{12(5):677--684},
  \href{https://doi.org/10.1109/TVCG.2006.160}{2006}.
  \href{https://doi.org/10.1109/TVCG.2006.160}
{doi: {{%
10\hspace{.1pt}\discretionary{.}{%
}{.}\hspace{.4pt}1109\discretionary{/}{%
}{/}TVCG\hspace{.1pt}\discretionary{.}{%
}{.}\hspace{.4pt}2006\hspace{.1pt}\discretionary{.}{%
}{.}\hspace{.4pt}160}}}


\bibitem{DBLP:journals/tvcg/HenryFM07}
\href{https://doi.org/10.1109/TVCG.2007.70582}{N.~Henry, J.~Fekete, and M.~J.
  McGuffin}.
\newblock \href{https://doi.org/10.1109/TVCG.2007.70582}{Nodetrix: a hybrid
  visualization of social networks}.
\newblock \href{https://doi.org/10.1109/TVCG.2007.70582}{{\em IEEE Trans. Vis.
  Comp. Graph.}},
  \href{https://doi.org/10.1109/TVCG.2007.70582}{13(6):1302--1309},
  \href{https://doi.org/10.1109/TVCG.2007.70582}{2007}.
  \href{https://doi.org/10.1109/TVCG.2007.70582}
{doi: {{%
10\hspace{.1pt}\discretionary{.}{%
}{.}\hspace{.4pt}1109\discretionary{/}{%
}{/}TVCG\hspace{.1pt}\discretionary{.}{%
}{.}\hspace{.4pt}2007\hspace{.1pt}\discretionary{.}{%
}{.}\hspace{.4pt}70582}}}


\bibitem{newshape}
\href{https://doi.org/https://doi.org/10.1109/PacificVis53943.2022.00014}{S.~Hong,
  A.~Meidiana, J.~Wood, J.~P. Ataides, P.~Eades, and K.~Park}.
\newblock
  \href{https://doi.org/https://doi.org/10.1109/PacificVis53943.2022.00014}{{dGG,
  dRNG, DSC:} new degree-based shape-based faithfulness metrics for large and
  complex graph visualization}.
\newblock
  \href{https://doi.org/https://doi.org/10.1109/PacificVis53943.2022.00014}{In
  {\em IEEE PacificVis 2022}},
  \href{https://doi.org/https://doi.org/10.1109/PacificVis53943.2022.00014}{pp.
  51--60},
  \href{https://doi.org/https://doi.org/10.1109/PacificVis53943.2022.00014}{2022}.
  \href{https://doi.org/10.1109/PacificVis53943.2022.00014}
{doi: {{%
10\hspace{.1pt}\discretionary{.}{%
}{.}\hspace{.4pt}1109\discretionary{/}{%
}{/}PacificVis53943\hspace{.1pt}\discretionary{.}{%
}{.}\hspace{.4pt}2022\hspace{.1pt}\discretionary{.}{%
}{.}\hspace{.4pt}00014}}}


\bibitem{Houtkamp2003}
\href{https://doi.org/10.3758/BF03194840}{R.~Houtkamp, H.~Spekreijse, and P.~R.
  Roelfsema}.
\newblock \href{https://doi.org/10.3758/BF03194840}{{A gradual spread of
  attention}}.
\newblock \href{https://doi.org/10.3758/BF03194840}{{\em Perception \&
  Psychophysics}}, \href{https://doi.org/10.3758/BF03194840}{65(7):1136--1144},
  \href{https://doi.org/10.3758/BF03194840}{2003}.
  \href{https://doi.org/10.3758/BF03194840}
{doi: {{%
10\hspace{.1pt}\discretionary{.}{%
}{.}\hspace{.4pt}3758\discretionary{/}{%
}{/}BF03194840}}}


\bibitem{Itti2005}
\href{https://doi.org/10.1016/B978-012375731-9/50098-7}{L.~Itti}.
\newblock \href{https://doi.org/10.1016/B978-012375731-9/50098-7}{{Models of
  Bottom-up Attention and Saliency}}.
\newblock \href{https://doi.org/10.1016/B978-012375731-9/50098-7}{In {\em
  Neurobiology of Attention}},
  \href{https://doi.org/10.1016/B978-012375731-9/50098-7}{pp. 576--582}.
  \href{https://doi.org/10.1016/B978-012375731-9/50098-7}{Elsevier},
  \href{https://doi.org/10.1016/B978-012375731-9/50098-7}{2005}.
  \href{https://doi.org/10.1016/B978-012375731-9/50098-7}
{doi: {{%
10\hspace{.1pt}\discretionary{.}{%
}{.}\hspace{.4pt}1016\discretionary{/}{%
}{/}B978\discretionary{%
}{-}{-}012375731\discretionary{%
}{-}{-}9\discretionary{/}{%
}{/}50098\discretionary{%
}{-}{-}7}}}


\bibitem{tilt}
\href{https://doi.org/10.1097/MAO.0b013e31821c6c7b}{M.~Janssen, M.~Lauvenberg,
  W.~{van der Ven}, T.~Bloebaum, and H.~Kingma}.
\newblock \href{https://doi.org/10.1097/MAO.0b013e31821c6c7b}{Perception
  threshold for tilt}.
\newblock \href{https://doi.org/10.1097/MAO.0b013e31821c6c7b}{{\em Otol
  Neurotol.}},
  \href{https://doi.org/10.1097/MAO.0b013e31821c6c7b}{32(5):818--825},
  \href{https://doi.org/10.1097/MAO.0b013e31821c6c7b}{2011}.
  \href{https://doi.org/10.1097/MAO.0b013e31821c6c7b}
{doi: {{%
10\hspace{.1pt}\discretionary{.}{%
}{.}\hspace{.4pt}1097\discretionary{/}{%
}{/}MAO\hspace{.1pt}\discretionary{.}{%
}{.}\hspace{.4pt}0b013e31821c6c7b}}}


\bibitem{DBLP:journals/ipl/KamadaK89}
\href{https://doi.org/10.1016/0020-0190(89)90102-6}{T.~Kamada and S.~Kawai}.
\newblock \href{https://doi.org/10.1016/0020-0190(89)90102-6}{An algorithm for
  drawing general undirected graphs}.
\newblock \href{https://doi.org/10.1016/0020-0190(89)90102-6}{{\em Inf.
  Process. Lett.}},
  \href{https://doi.org/10.1016/0020-0190(89)90102-6}{31(1):7--15},
  \href{https://doi.org/10.1016/0020-0190(89)90102-6}{1989}.
  \href{https://doi.org/10.1016/0020-0190(89)90102-6}
{doi: {{%
10\hspace{.1pt}\discretionary{.}{%
}{.}\hspace{.4pt}1016\discretionary{/}{%
}{/}0020\discretionary{%
}{-}{-}0190\discretionary{%
}{(}{(}89\discretionary{)}{%
}{)}90102\discretionary{%
}{-}{-}6}}}


\bibitem{kuratowski}
C.~Kuratowski.
\newblock Sur le problème des courbes gauches en topologie.
\newblock {\em Fundamenta Mathematicae}, 15:271--283, 1930.

\bibitem{DBLP:conf/gd/LazardLL17}
\href{https://doi.org/10.1007/978-3-319-73915-1_2}{S.~Lazard, W.~J. Lenhart,
  and G.~Liotta}.
\newblock \href{https://doi.org/10.1007/978-3-319-73915-1_2}{On the edge-length
  ratio of outerplanar graphs}.
\newblock \href{https://doi.org/10.1007/978-3-319-73915-1_2}{In F.~Frati and
  K.~Ma, eds., {\em Graph Drawing and Network Visualization - 25th
  International Symposium, {GD} 2017, Boston, MA, USA, September 25-27, 2017,
  Revised Selected Papers}},
  \href{https://doi.org/10.1007/978-3-319-73915-1_2}{vol. 10692 of {\em Lecture
  Notes in Computer Science}},
  \href{https://doi.org/10.1007/978-3-319-73915-1_2}{pp. 17--23}.
  \href{https://doi.org/10.1007/978-3-319-73915-1_2}{Springer},
  \href{https://doi.org/10.1007/978-3-319-73915-1_2}{2017}.
  \href{https://doi.org/10.1007/978-3-319-73915-1_2}
{doi: {{%
10\hspace{.1pt}\discretionary{.}{%
}{.}\hspace{.4pt}1007\discretionary{/}{%
}{/}978\discretionary{%
}{-}{-}3\discretionary{%
}{-}{-}319\discretionary{%
}{-}{-}73915\discretionary{%
}{-}{-}1\_2}}}


\bibitem{DBLP:journals/tcs/LazardLL19}
\href{https://doi.org/10.1016/j.tcs.2018.10.002}{S.~Lazard, W.~J. Lenhart, and
  G.~Liotta}.
\newblock \href{https://doi.org/10.1016/j.tcs.2018.10.002}{On the edge-length
  ratio of outerplanar graphs}.
\newblock \href{https://doi.org/10.1016/j.tcs.2018.10.002}{{\em Theor. Comput.
  Sci.}}, \href{https://doi.org/10.1016/j.tcs.2018.10.002}{770:88--94},
  \href{https://doi.org/10.1016/j.tcs.2018.10.002}{2019}.
  \href{https://doi.org/10.1016/j.tcs.2018.10.002}
{doi: {{%
10\hspace{.1pt}\discretionary{.}{%
}{.}\hspace{.4pt}1016\discretionary{/}{%
}{/}j\hspace{.1pt}\discretionary{.}{%
}{.}\hspace{.4pt}tcs\hspace{.1pt}\discretionary{.}{%
}{.}\hspace{.4pt}2018\hspace{.1pt}\discretionary{.}{%
}{.}\hspace{.4pt}10\hspace{.1pt}\discretionary{.}{%
}{.}\hspace{.4pt}002}}}


\bibitem{Luck2008}
\href{https://doi.org/10.1093/acprof:oso/9780195305487.001.0001}{S.~J. Luck and
  A.~R. Hollingworth}.
\newblock \href{https://doi.org/10.1093/acprof:oso/9780195305487.001.0001}{{\em
  {Visual Memory}}}.
\newblock
  \href{https://doi.org/10.1093/acprof:oso/9780195305487.001.0001}{Oxford
  University Press US},
  \href{https://doi.org/10.1093/acprof:oso/9780195305487.001.0001}{2008}.
  \href{https://doi.org/10.1093/acprof:oso/9780195305487.001.0001}
{doi: {{%
10\hspace{.1pt}\discretionary{.}{%
}{.}\hspace{.4pt}1093\discretionary{/}{%
}{/}acprof\discretionary{:}{%
}{:}oso\discretionary{/}{%
}{/}9780195305487\hspace{.1pt}\discretionary{.}{%
}{.}\hspace{.4pt}001\hspace{.1pt}\discretionary{.}{%
}{.}\hspace{.4pt}0001}}}


\bibitem{DBLP:journals/csr/McConnellMNS11}
\href{https://doi.org/10.1016/j.cosrev.2010.09.009}{R.~M. McConnell,
  K.~Mehlhorn, S.~N{\"{a}}her, and P.~Schweitzer}.
\newblock \href{https://doi.org/10.1016/j.cosrev.2010.09.009}{Certifying
  algorithms}.
\newblock \href{https://doi.org/10.1016/j.cosrev.2010.09.009}{{\em Comput. Sci.
  Review}}, \href{https://doi.org/10.1016/j.cosrev.2010.09.009}{5(2):119--161},
  \href{https://doi.org/10.1016/j.cosrev.2010.09.009}{2011}.
  \href{https://doi.org/10.1016/j.cosrev.2010.09.009}
{doi: {{%
10\hspace{.1pt}\discretionary{.}{%
}{.}\hspace{.4pt}1016\discretionary{/}{%
}{/}j\hspace{.1pt}\discretionary{.}{%
}{.}\hspace{.4pt}cosrev\hspace{.1pt}\discretionary{.}{%
}{.}\hspace{.4pt}2010\hspace{.1pt}\discretionary{.}{%
}{.}\hspace{.4pt}09\hspace{.1pt}\discretionary{.}{%
}{.}\hspace{.4pt}009}}}


\bibitem{change}
\href{https://doi.org/10.1007/978-3-030-68766-3_35}{A.~Meidiana, S.~Hong, and
  P.~Eades}.
\newblock \href{https://doi.org/10.1007/978-3-030-68766-3_35}{New quality
  metrics for dynamic graph drawing}.
\newblock \href{https://doi.org/10.1007/978-3-030-68766-3_35}{In D.~Auber and
  P.~Valtr, eds., {\em Graph Drawing and Network Visualization}},
  \href{https://doi.org/10.1007/978-3-030-68766-3_35}{pp. 450--465},
  \href{https://doi.org/10.1007/978-3-030-68766-3_35}{2020}.
  \href{https://doi.org/10.1007/978-3-030-68766-3_35}
{doi: {{%
10\hspace{.1pt}\discretionary{.}{%
}{.}\hspace{.4pt}1007\discretionary{/}{%
}{/}978\discretionary{%
}{-}{-}3\discretionary{%
}{-}{-}030\discretionary{%
}{-}{-}68766\discretionary{%
}{-}{-}3\_35}}}


\bibitem{cluster}
\href{https://doi.org/10.1007/978-3-030-35802-0_10}{A.~Meidiana, S.~Hong,
  P.~Eades, and D.~A. Keim}.
\newblock \href{https://doi.org/10.1007/978-3-030-35802-0_10}{A quality metric
  for visualization of clusters in graphs}.
\newblock \href{https://doi.org/10.1007/978-3-030-35802-0_10}{In D.~Archambault
  and C.~D. T{\'{o}}th, eds., {\em Graph Drawing and Network Visualization}},
  \href{https://doi.org/10.1007/978-3-030-35802-0_10}{pp. 125--138},
  \href{https://doi.org/10.1007/978-3-030-35802-0_10}{2019}.
  \href{https://doi.org/10.1007/978-3-030-35802-0_10}
{doi: {{%
10\hspace{.1pt}\discretionary{.}{%
}{.}\hspace{.4pt}1007\discretionary{/}{%
}{/}978\discretionary{%
}{-}{-}3\discretionary{%
}{-}{-}030\discretionary{%
}{-}{-}35802\discretionary{%
}{-}{-}0\_10}}}


\bibitem{symmetry}
\href{https://doi.org/10.1109/PACIFICVIS48177.2020.1022}{A.~Meidiana, S.~Hong,
  P.~Eades, and D.~A. Keim}.
\newblock \href{https://doi.org/10.1109/PACIFICVIS48177.2020.1022}{Quality
  metrics for symmetric graph drawings}.
\newblock \href{https://doi.org/10.1109/PACIFICVIS48177.2020.1022}{In {\em IEEE
  PacificVis 2020}},
  \href{https://doi.org/10.1109/PACIFICVIS48177.2020.1022}{pp. 11--15},
  \href{https://doi.org/10.1109/PACIFICVIS48177.2020.1022}{2020}.
  \href{https://doi.org/10.1109/PACIFICVIS48177.2020.1022}
{doi: {{%
10\hspace{.1pt}\discretionary{.}{%
}{.}\hspace{.4pt}1109\discretionary{/}{%
}{/}PACIFICVIS48177\hspace{.1pt}\discretionary{.}{%
}{.}\hspace{.4pt}2020\hspace{.1pt}\discretionary{.}{%
}{.}\hspace{.4pt}1022}}}


\bibitem{Miyake2012}
\href{https://doi.org/10.1177/0963721411429458}{A.~Miyake and N.~P. Friedman}.
\newblock \href{https://doi.org/10.1177/0963721411429458}{{The nature and
  organization of individual differences in executive functions: Four general
  conclusions}}.
\newblock \href{https://doi.org/10.1177/0963721411429458}{{\em Cur. Direct.
  Psychol. Sci.}},
  \href{https://doi.org/10.1177/0963721411429458}{21(1):8--14},
  \href{https://doi.org/10.1177/0963721411429458}{2012}.
  \href{https://doi.org/10.1177/0963721411429458}
{doi: {{%
10\hspace{.1pt}\discretionary{.}{%
}{.}\hspace{.4pt}1177\discretionary{/}{%
}{/}0963721411429458}}}


\bibitem{NI}
\href{https://doi.org/10.1007/BF01758778}{H.~Nagamochi and T.~Ibaraki}.
\newblock \href{https://doi.org/10.1007/BF01758778}{A linear-time algorithm for
  finding a sparse k-connected spanning subgraph of a k-connected graph}.
\newblock \href{https://doi.org/10.1007/BF01758778}{{\em Algorithmica}},
  \href{https://doi.org/10.1007/BF01758778}{7(5{\&}6):583--596},
  \href{https://doi.org/10.1007/BF01758778}{1992}.
  \href{https://doi.org/10.1007/BF01758778}
{doi: {{%
10\hspace{.1pt}\discretionary{.}{%
}{.}\hspace{.4pt}1007\discretionary{/}{%
}{/}BF01758778}}}


\bibitem{nelsen1993proofs}
\href{https://books.google.de/books?id=Kx2cjyzTIYkC}{R.~Nelsen}.
\newblock \href{https://books.google.de/books?id=Kx2cjyzTIYkC}{{\em Proofs
  Without Words: Exercises in Visual Thinking}}.
\newblock \href{https://books.google.de/books?id=Kx2cjyzTIYkC}{Classroom
  resource materials}.
  \href{https://books.google.de/books?id=Kx2cjyzTIYkC}{Mathematical Association
  of America}, \href{https://books.google.de/books?id=Kx2cjyzTIYkC}{1993}.

\bibitem{faithful}
\href{https://doi.org/10.1109/PACIFICVIS.2013.6596147}{Q.~H. Nguyen, P.~Eades,
  and S.~Hong}.
\newblock \href{https://doi.org/10.1109/PACIFICVIS.2013.6596147}{On the
  faithfulness of graph visualizations}.
\newblock \href{https://doi.org/10.1109/PACIFICVIS.2013.6596147}{In
  S.~Carpendale, W.~Chen, and S.~Hong, eds., {\em {IEEE} PacificVis}},
  \href{https://doi.org/10.1109/PACIFICVIS.2013.6596147}{pp. 209--216},
  \href{https://doi.org/10.1109/PACIFICVIS.2013.6596147}{2013}.
  \href{https://doi.org/10.1109/PACIFICVIS.2013.6596147}
{doi: {{%
10\hspace{.1pt}\discretionary{.}{%
}{.}\hspace{.4pt}1109\discretionary{/}{%
}{/}PACIFICVIS\hspace{.1pt}\discretionary{.}{%
}{.}\hspace{.4pt}2013\hspace{.1pt}\discretionary{.}{%
}{.}\hspace{.4pt}6596147}}}


\bibitem{DBLP:books/ws/NishizekiR04}
\href{https://doi.org/10.1142/5648}{T.~Nishizeki and M.~S. Rahman}.
\newblock \href{https://doi.org/10.1142/5648}{{\em Planar Graph Drawing}},
  \href{https://doi.org/10.1142/5648}{vol.~12 of {\em Lect. Notes Ser.
  Computing}}.
\newblock \href{https://doi.org/10.1142/5648}{World Scientific},
  \href{https://doi.org/10.1142/5648}{2004}.
  \href{https://doi.org/10.1142/5648}
{doi: {{%
10\hspace{.1pt}\discretionary{.}{%
}{.}\hspace{.4pt}1142\discretionary{/}{%
}{/}5648}}}


\bibitem{oxford}
\href{https://doi.org/10.1093/OED/3549282820}{{Oxford English Dictionary}}.
\newblock \href{https://doi.org/10.1093/OED/3549282820}{four-colour |
  four-color, adj.}
\newblock \href{https://doi.org/10.1093/OED/3549282820}{In {\em Oxford English
  Dictionary}}. \href{https://doi.org/10.1093/OED/3549282820}{Oxford University
  Press}, \href{https://doi.org/10.1093/OED/3549282820}{2023}.

\bibitem{pirolli2005sensemaking}
P.~Pirolli and S.~Card.
\newblock The sensemaking process and leverage points for analyst technology as
  identified through cognitive task analysis.
\newblock In {\em Proc.~Intl.~Conf.~Intelligence Analysis}, vol.~5, pp. 2--4.
  McLean, VA, USA, 2005.

\bibitem{Purchase2002}
\href{https://doi.org/10.1016/S1045-926X(02)90232-6}{H.~Purchase}.
\newblock \href{https://doi.org/10.1016/S1045-926X(02)90232-6}{Metrics for
  graph drawing aesthetics}.
\newblock \href{https://doi.org/10.1016/S1045-926X(02)90232-6}{{\em J. Vis.
  Lang. \& Comput.}},
  \href{https://doi.org/10.1016/S1045-926X(02)90232-6}{13(5):501--516},
  \href{https://doi.org/10.1016/S1045-926X(02)90232-6}{2002}.
  \href{https://doi.org/10.1016/S1045-926X(02)90232-6}
{doi: {{%
10\hspace{.1pt}\discretionary{.}{%
}{.}\hspace{.4pt}1016\discretionary{/}{%
}{/}S1045\discretionary{%
}{-}{-}926X\discretionary{%
}{(}{(}02\discretionary{)}{%
}{)}90232\discretionary{%
}{-}{-}6}}}


\bibitem{knowledgegen}
D.~Sacha, A.~Stoffel, F.~Stoffel, B.~C. Kwon, G.~Ellis, and D.~A. Keim.
\newblock Knowledge generation model for visual analytics.
\newblock {\em IEEE Trans.~Vis.~Comp.~ Graph.}, 20(12):1604--1613, 2014.
\newblock 10.1109/TVCG.2014.2346481.

\bibitem{ST99}
\href{https://doi.org/10.1007/3-540-46648-7_11}{J.~M. Six and I.~G. Tollis}.
\newblock \href{https://doi.org/10.1007/3-540-46648-7_11}{A framework for
  circular drawings of networks}.
\newblock \href{https://doi.org/10.1007/3-540-46648-7_11}{In
  J.~Kratochv{\'{\i}}l, ed., {\em Proc.~7th Intl.~Symp.~Graph Drawing}},
  \href{https://doi.org/10.1007/3-540-46648-7_11}{vol. 1731 of {\em LNCS}},
  \href{https://doi.org/10.1007/3-540-46648-7_11}{pp. 107--116}.
  \href{https://doi.org/10.1007/3-540-46648-7_11}{Springer},
  \href{https://doi.org/10.1007/3-540-46648-7_11}{1999}.
  \href{https://doi.org/10.1007/3-540-46648-7_11}
{doi: {{%
10\hspace{.1pt}\discretionary{.}{%
}{.}\hspace{.4pt}1007\discretionary{/}{%
}{/}3\discretionary{%
}{-}{-}540\discretionary{%
}{-}{-}46648\discretionary{%
}{-}{-}7\_11}}}


\bibitem{9966829}
\href{https://doi.org/10.1109/TVCG.2022.3225554}{S.~Song, C.~Li, D.~Li,
  J.~Chen, and C.~Wang}.
\newblock \href{https://doi.org/10.1109/TVCG.2022.3225554}{Graphdecoder:
  Recovering diverse network graphs from visualization images via
  attention-aware learning}.
\newblock \href{https://doi.org/10.1109/TVCG.2022.3225554}{{\em IEEE
  Trans.~Vis.~Comp.~Graph.}},
  \href{https://doi.org/10.1109/TVCG.2022.3225554}{pp. 1--17},
  \href{https://doi.org/10.1109/TVCG.2022.3225554}{2022}.
  \href{https://doi.org/10.1109/TVCG.2022.3225554}
{doi: {{%
10\hspace{.1pt}\discretionary{.}{%
}{.}\hspace{.4pt}1109\discretionary{/}{%
}{/}TVCG\hspace{.1pt}\discretionary{.}{%
}{.}\hspace{.4pt}2022\hspace{.1pt}\discretionary{.}{%
}{.}\hspace{.4pt}3225554}}}


\bibitem{DBLP:reference/crc/2013gd}
\href{https://www.crcpress.com/Handbook-of-Graph-Drawing-and-Visualization/Tamassia/9781584884125}{R.~Tamassia,
  ed.}
\newblock
  \href{https://www.crcpress.com/Handbook-of-Graph-Drawing-and-Visualization/Tamassia/9781584884125}{{\em
  Handbook on Graph Drawing and Visualization}}.
\newblock
  \href{https://www.crcpress.com/Handbook-of-Graph-Drawing-and-Visualization/Tamassia/9781584884125}{Chapman
  and Hall},
  \href{https://www.crcpress.com/Handbook-of-Graph-Drawing-and-Visualization/Tamassia/9781584884125}{2013}.

\bibitem{DBLP:books/daglib/0001351}
E.~R. Tufte.
\newblock {\em The visual display of quantitative information}.
\newblock Graphics Press, 1992.

\bibitem{DBLP:books/lib/Tufte97}
\href{https://www.worldcat.org/oclc/36234417}{E.~R. Tufte}.
\newblock \href{https://www.worldcat.org/oclc/36234417}{{\em Visual
  explanations - images and quantities, evidence and narrative}}.
\newblock \href{https://www.worldcat.org/oclc/36234417}{Graphics Press},
  \href{https://www.worldcat.org/oclc/36234417}{1997}.

\bibitem{Tufte2001}
E.~R. Tufte.
\newblock {\em {The Visual Display of Quantitative Information}}.
\newblock Graphics Press, 2001.

\bibitem{philosophicalImplications}
T.~Tymoczko.
\newblock The four-color problem and its philosophical significance.
\newblock {\em J. Philosophy}, 76(2):57--83, 1979.

\bibitem{art}
B.~Victor.
\newblock Scientific communication as sequential art.
\newblock \url{http://worrydream.com/ScientificCommunicationAsSequentialArt/},
  2011.

\bibitem{Wagemans2012}
\href{https://doi.org/10.1037/a0029333}{J.~Wagemans, J.~H. Elder, M.~Kubovy,
  S.~E. Palmer, M.~A. Peterson, M.~Singh, and R.~von~der Heydt}.
\newblock \href{https://doi.org/10.1037/a0029333}{A century of gestalt
  psychology in visual perception: I. perceptual grouping and figure–ground
  organization}.
\newblock \href{https://doi.org/10.1037/a0029333}{{\em Psychol. Bulletin}},
  \href{https://doi.org/10.1037/a0029333}{138(6):1172--1217},
  \href{https://doi.org/10.1037/a0029333}{2012}.
  \href{https://doi.org/10.1037/a0029333}
{doi: {{%
10\hspace{.1pt}\discretionary{.}{%
}{.}\hspace{.4pt}1037\discretionary{/}{%
}{/}a0029333}}}


\bibitem{Ware}
C.~Ware.
\newblock {\em Information Visualization – Perception for Design}.
\newblock Morgan Kaufmann, 2004.

\bibitem{Ware2021}
\href{https://doi.org/https://doi.org/10.1016/C2016-0-01395-5}{C.~Ware}.
\newblock \href{https://doi.org/https://doi.org/10.1016/C2016-0-01395-5}{{\em
  {Visual Thinking For Information Design}}}.
\newblock \href{https://doi.org/https://doi.org/10.1016/C2016-0-01395-5}{Morgan
  Kaufmann},
  \href{https://doi.org/https://doi.org/10.1016/C2016-0-01395-5}{2021}.
  \href{https://doi.org/10.1016/C2016-0-01395-5}
{doi: {{%
10\hspace{.1pt}\discretionary{.}{%
}{.}\hspace{.4pt}1016\discretionary{/}{%
}{/}C2016\discretionary{%
}{-}{-}0\discretionary{%
}{-}{-}01395\discretionary{%
}{-}{-}5}}}


\bibitem{WS}
\href{https://doi.org/https://doi.org/10.1017/CBO9780511815478}{S.~Wasserman
  and K.~Faust}.
\newblock \href{https://doi.org/https://doi.org/10.1017/CBO9780511815478}{{\em
  Social network analysis: Methods and applications}}.
\newblock
  \href{https://doi.org/https://doi.org/10.1017/CBO9780511815478}{Cambridge
  University Press},
  \href{https://doi.org/https://doi.org/10.1017/CBO9780511815478}{1994}.
  \href{https://doi.org/10.1017/CBO9780511815478}
{doi: {{%
10\hspace{.1pt}\discretionary{.}{%
}{.}\hspace{.4pt}1017\discretionary{/}{%
}{/}CBO9780511815478}}}


\bibitem{nature}
\href{https://doi.org/10.1038/30918}{D.~J. Watts and S.~H. Strogatz}.
\newblock \href{https://doi.org/10.1038/30918}{Collective dynamics of
  ‘small-world’ networks}.
\newblock \href{https://doi.org/10.1038/30918}{{\em Nature}},
  \href{https://doi.org/10.1038/30918}{393:440--442},
  \href{https://doi.org/10.1038/30918}{1998}.
  \href{https://doi.org/10.1038/30918}
{doi: {{%
10\hspace{.1pt}\discretionary{.}{%
}{.}\hspace{.4pt}1038\discretionary{/}{%
}{/}30918}}}


\bibitem{DBLP:journals/tvcg/WickhamCHB10}
\href{https://doi.org/10.1109/TVCG.2010.161}{H.~Wickham, D.~Cook, H.~Hofmann,
  and A.~Buja}.
\newblock \href{https://doi.org/10.1109/TVCG.2010.161}{Graphical inference for
  {InfoVis}}.
\newblock \href{https://doi.org/10.1109/TVCG.2010.161}{{\em {IEEE}
  Trans.~Vis.~Comp.~Graph.}},
  \href{https://doi.org/10.1109/TVCG.2010.161}{16(6):973--979},
  \href{https://doi.org/10.1109/TVCG.2010.161}{2010}.
  \href{https://doi.org/10.1109/TVCG.2010.161}
{doi: {{%
10\hspace{.1pt}\discretionary{.}{%
}{.}\hspace{.4pt}1109\discretionary{/}{%
}{/}TVCG\hspace{.1pt}\discretionary{.}{%
}{.}\hspace{.4pt}2010\hspace{.1pt}\discretionary{.}{%
}{.}\hspace{.4pt}161}}}


\bibitem{wiki}
Wikipedia.
\newblock Mathematical proof.
\newblock \url{https://en.wikipedia.org/wiki/Mathematical_proof##Visual_proof},
  2023.

\bibitem{Wolfe2010}
\href{https://doi.org/10.1016/j.cub.2010.02.016}{J.~Wolfe}.
\newblock \href{https://doi.org/10.1016/j.cub.2010.02.016}{{Visual search}}.
\newblock \href{https://doi.org/10.1016/j.cub.2010.02.016}{{\em Cur. Biol.}},
  \href{https://doi.org/10.1016/j.cub.2010.02.016}{20(8):R346--R349},
  \href{https://doi.org/10.1016/j.cub.2010.02.016}{2010}.
  \href{https://doi.org/10.1016/j.cub.2010.02.016}
{doi: {{%
10\hspace{.1pt}\discretionary{.}{%
}{.}\hspace{.4pt}1016\discretionary{/}{%
}{/}j\hspace{.1pt}\discretionary{.}{%
}{.}\hspace{.4pt}cub\hspace{.1pt}\discretionary{.}{%
}{.}\hspace{.4pt}2010\hspace{.1pt}\discretionary{.}{%
}{.}\hspace{.4pt}02\hspace{.1pt}\discretionary{.}{%
}{.}\hspace{.4pt}016}}}


\bibitem{Yan89}
\href{https://doi.org/10.1016/0022-0000(89)90032-9}{M.~Yannakakis}.
\newblock \href{https://doi.org/10.1016/0022-0000(89)90032-9}{Embedding planar
  graphs in four pages}.
\newblock \href{https://doi.org/10.1016/0022-0000(89)90032-9}{{\em J. Comput.
  Syst. Sci.}},
  \href{https://doi.org/10.1016/0022-0000(89)90032-9}{38(1):36--67},
  \href{https://doi.org/10.1016/0022-0000(89)90032-9}{1989}.
  \href{https://doi.org/10.1016/0022-0000(89)90032-9}
{doi: {{%
10\hspace{.1pt}\discretionary{.}{%
}{.}\hspace{.4pt}1016\discretionary{/}{%
}{/}0022\discretionary{%
}{-}{-}0000\discretionary{%
}{(}{(}89\discretionary{)}{%
}{)}90032\discretionary{%
}{-}{-}9}}}


\bibitem{yed}
yWorks.
\newblock y{E}d - graph editor.
\newblock \url{https://www.yworks.com/products/yed}, 2023.

\end{thebibliography}

\newpage

\appendix

\section{Ommited Material from Section~\ref{sec:preliminaries}}
\label{app:preliminaries}

\subsection{Examples for Certifying Algorithms}
\label{app:certifying}
As an example, 
reconsider~\cref{ex:mst}. Here, our function $f$ takes as an input a graph $G$ and should output either \texttt{true} if the graph is connected or \texttt{false} otherwise. A witness for a positive instance would be a spanning tree $T$ as discussed in~\cref{sec:connected}. Given $T$, we can easily see that $G$ is connected and we can determine this by checking that every vertex is part of $T$; i.\,e.,~Property~\ref{prop:certifying:3} holds. In fact, this checking can be done efficiently by running a BFS on $T$, that is, the time for checking the correctness of the certificate depends only on the number of vertices, not on the number of edges; i.\,e.,~Property~\ref{prop:certifying:2} holds. 
On the other hand, if $G$ is not connected, the certifying algorithm for $f$ can provide a partition of the vertices into two disconnected sets $V_1$ and $V_2$ as a witness. We may check that we cannot reach any vertex in $V_2$ if we run a BFS starting from a vertex in $V_1$ to establish that $f$ was computed correctly; i.\,e.,~Property~\ref{prop:certifying:3} holds. If it is the case that $|V_1| \leq |V_2|$ we end up considering at most half of the vertices in this process, hence we may argue that Property~\ref{prop:certifying:2} holds. 
Since we found a good witness for positive and negative instances, we conclude that Property~\ref{prop:certifying:1} holds. 

While Ex.~\ref{ex:cut-vertex} also admits a certifying algorithm, 
%
for Ex.~\ref{ex:hamiltonian}, two issues emerge: Computing a Hamiltonian cycle is NP-complete, hence, the algorithm for $f$ may not run in polynomial time. However, even if we ignore this, we end up with the problem, that for a negative instance no easy witness is known, i.\,e., we do not even know how to guarantee Property~\ref{prop:certifying:1}. This appears to be a general issue for NP-complete assertions as a witness for a negative instance would be a short certificate for a CoNP-complete problem.

\section{Omitted Material from Section~\ref{sec:model}}
\label{app:model}

\subsection{Application of Visual Certificate and Provability Properties to Sections~\ref{ex:cut-vertex}, \ref{ex:mst} and \ref{ex:hamiltonian}}
\label{app:certificate}

Reconsider Ex.~\ref{ex:cut-vertex} to~\ref{ex:hamiltonian}. All drawings shown in~\cref{sec:examples:easy} have been \emph{unimpeachable} (Property~\ref{prop:certificate:1}), possibly with a single exception: In the drawing in~\cref{fig:cutvertex:force-directed}, one may argue that it is impossible for the judge to perceive whether there is an edge hidden behind the alleged cut-vertex questioning the unimpeachability of the layout. While in~\cref{fig:cutvertex:proof} it is even more difficult to follow all edges, it makes it easy to perceive all parts of the graph necessary for the validation. 

On the other hand, the drawing in~\cref{fig:cutvertex:force-directed} also fails the \emph{simplicity} requirement (Property~\ref{prop:certificate:3}). Namely, one could actually hide an edge behind the cut-vertex that connects the alleged left and right component. Most likely, $\mathcal{M}(G)$ would consist of two connected components connected at a single vertex, as the drawing shows two vaguely compact salient features touching at a single point. Hence, if we showed the judge two drawings of that type where in one of the drawings an edge was hidden behind the cut-vertex, the judge would be unable to \emph{distinguish} the correct certificate from the wrong one. In contrast, the drawing in~\cref{fig:cutvertex:proof} circumvents this problem, as the cut-vertex is the only vertex at the bottom of the drawing. Similarly, the highlighted parts in~\cref{fig:st:middle,fig:st:good,fig:hamilton} draw the judge's attention towards them, hence, these structures would become part of the mental model. Given the highlighted parts, it is easy to prove connectivity or the existence of the Hamiltonian cycle.

Finally, for the drawings in~\cref{fig:st:middle,fig:hamilton1} it is  not trivial to \emph{check} whether the assertion is correct (Property~\ref{prop:certificate:2}). The mental model $\mathcal{M}(G)$ will consist of a tangle of highlighted edges embedded in an even larger tangle of edges. While the judge will most likely succeed to establish that the highlighted parts of the graph form a tree or a cycle, they might have to perform a BFS. Hence, especially in the case of~\cref{ex:mst}, this would be as efficient as computing the validity of the assertion from scratch and these visualizations are not to be considered visual proofs. On the other hand, the visualizations in~\cref{fig:st:good,fig:hamilton2,fig:hamilton3} lead to formation of mental models that highlight the tree or cycle in few components of the mental model. Hence the judge only needs to check if the mental model includes any vertices not belonging to the tree or  cycle.
Moreover, in~\cref{ex:hamiltonian} the judge also has to check that the salient feature representing the cycle in fact contains all edges. This can be easily done in~\cref{fig:hamilton2,fig:hamilton3}, however, if the size of the graph was larger, the process would be more efficient for the visualization in~\cref{fig:hamilton3} where it is sufficient to check that the long diagonal and the two single cells are present. On the other hand, for a visualization in the style of~\cref{fig:hamilton2}, it is required to assert the existence of each presumed edge along the cycle. Thus, more components of the mental model must be checked and the {perceptual complexity}, that is, the time that the judge needs to check the assertion given the mental model, depends on the size of the graph.

\subsection{Unimpeachability and the Defense Lawyer}
\label{sec:defense}

Recall that the task of the defense lawyer is to establish \emph{unimpeachability} (Property~\ref{prop:certificate:1}), i.\,e.,
we require that the visualization should display the ground truth properties of the graph, and that the judge can clearly perceive the parts of the visualization required for validating the assertion. The latter property is required so that the judge can extract the evidence embedded in the visual certificate whereas the former property ensures that a {non-certificate} which showcases evidence embedded in a visualization of a graph that is not identical to the input can be detected as such.
In the literature, these concepts are known as \emph{information faithfulness} and \emph{task readability}, respectively~\cite{faithful}. 
\emph{Faithfulness} refers to whether a visualization of a graph displays its ground truth properties and structures in a logically consistent manner, and the  \emph{readability} refers to the perceptual and cognitive interpretation of the visualization by the viewer~\cite{faithful}. 

More specifically, {\em information faithfulness} means that all the information about a graph $G$ is displayed in the visualization.
Faithfulness metrics are defined based on the type of ground truth structures of graphs, such as  {\em shape}~\cite{newshape},  \emph{cluster}~\cite{cluster},  {\em symmetry}~\cite{symmetry} and {\em change}~\cite{change} and can be appropriately selected depending on the application scenario. 
Similarly,  {\em task readability} means that the user can perceive enough information from the visualization to correctly perform the task, here validating the assertion. 
In our examples for good visual certificates, task readability was improved at the cost of information readability compared to standard layouts; see~\cref{tab:metrics}. 



Our definition of \emph{unimpeachability} is purposely quite objective so that it can be discussed efficiently when establishing new visual proof techniques. A purely logically acting defense lawyer should have no reason to raise doubts about the visual certificate if we obey these requirements and no feedback between defense lawyer and judge is necessary (hence the dashed connection in Fig.~\ref{fig:teaser}). 

\subsection{The Mental Model}
\label{sec:mental-model}

The mental model $\mathcal{M}(G)$ formed by the judge is an important component in assessing the usefulness of a visual certificate. Since~cognition \cite{Miyake2012} and perception~\cite{Grzeczkowski2017} differ from user to user, we must \emph{predict} the \emph{expected} mental model  $\mathcal{M}(G)$ instead of assessing the mental model of a specific user.
Hence, to influence the mental model, we have to carefully design visual certificates so to exploit known features of visual perception: In our analysis of~\cref{ex:cut-vertex} in~\cref{sec:cutvertex}, we relied on the effect of salient features being automatically grouped and perceived  as cohesive components~\cite{Wagemans2012}. Thus, we have good reason to assume that a human observer would see two components glued together at a vertex located below both components. In~\cref{sec:connected,sec:hamilton}, we exploited the fact that salient red-colored components naturally draw the attention of the user towards them~\cite{Itti2005}, so that they form a distinct shape in the foreground with the rest of the graph in the background. Thus, the graph layout guides the judge's perception and simplifies the analysis.

Another important aspect regarding the judges mental model is that we consider the judge to make an objective judgment based on the evidence encoded in the mental model. Thus, in our model the judge is not influenced by any prior knowledge or hypotheses regarding the data but at the same time will only accept the assertion if it has become irrefutable.   

\section{Visual Proofs for Additional Graph Properties}
\label{app:more-examples}

\begin{table}[t]
    \caption{Overview of graph properties with visual proofs presented \cref{app:more-examples}, and corresponding computational complexity of underlying properties as well as perceptual complexity of presented visual proofs ($n$ and $m$ denote the numbers of vertices and edges, resp.). 
    Shaded cells indicate open problems.}
    \centering
    \begin{tabular}{cccc}
        \toprule
         \cellcolor{blue!25} & \multicolumn{2}{c}{\cellcolor{blue!25}\bf Complexity} & \cellcolor{blue!25}\\
          \multirow{-2}{*}{\cellcolor{blue!25}\bf Assertion} &\cellcolor{blue!25} Computational & \cellcolor{blue!25}Perceptual & \multirow{-2}{*}{\cellcolor{blue!25}\bf Section}\\
          \aboverulesepcolor{blue!25}
         \midrule
         Queue number of $G$ is $\le k$   & NP-comp. & \cellcolor{blue!25} & \ref{sec:examples:canonical} \\
         Stack number of $G$ is $\le k$  & NP-comp. & \multirow{-2}{*}{\cellcolor{blue!25} $O(n$+$m)$} & \ref{sec:examples:canonical} \\
         \midrule
         $G$ is (outer-)planar & $O(n)$ & $O(n)$ \cellcolor{blue!25} & \ref{sec:examples:canonical} \\
         $G$ is \emph{not} (outer-)planar & $O(n)$ & $O(n)$ \cellcolor{blue!25} & \ref{sec:example:hard} \\
         \midrule
         $G$ is $k$-connected  & {$O(k^3 n^2)$} & $O(kn)$ \cellcolor{blue!25} & \ref{sec:example:hard} \\
         $G$ contains a size-$k$ clique & NP-comp. & $O(k)$ & \ref{sec:example:hard} \\
         $G$ has a size-$k$ independent set  & NP-comp. & $O(k)$ & \ref{sec:example:hard}   \\
         $G$ has a size-$k$ dominating set  & NP-comp. & $O(n)$ & \ref{sec:example:hard} \\
         \midrule
         
         Diameter of $G$ is $>k$   & $O(nm)$ & $O(k)$ & \ref{sec:examples:network-analysis} \\
         $d(u,v) = k$ & $O(n$+$m)$ & $O(k)$ & \ref{sec:examples:network-analysis} \\
         Diameter of $G$ is $\leq k$  & $O(nm)$ & \cellcolor{blue!25}  & \ref{sec:examples:network-analysis} \\
         Center of a graph  & $O(nm)$ & \multirow{-2}{*}{\cellcolor{blue!25} $O(n$+$m)$} & \ref{sec:examples:network-analysis} \\
         \midrule
         centrality, e.g., betweenness   & $O(nm)$ & \cellcolor{blue!25}  & \ref{sec:examples:network-analysis} \\
         $k$-core analysis &  $O(n$+$m)$ & \cellcolor{blue!25} & \ref{sec:examples:network-analysis} \\
         Automorphic equivalence   &  Isomorp.-c. & \multirow{-3}{*}{\cellcolor{blue!25} $O(n$+$m)$} & \ref{sec:examples:network-analysis} \\
         \bottomrule
    \end{tabular}
    \label{tab:overview:2}
\end{table}

In this appendix, we consider domain-specific properties from graph drawing and network analysis as well as assertions which are difficult to visually prove. For an overview of the results, see \cref{tab:overview:2}.

\subsection{Visual Proofs using Canonical Representations}
\label{sec:examples:canonical}

We discuss some graph properties that have historically been associated with certain representations that can serve as visual certificates.

\subparagraph{Planarity and Outer-Planarity.} Planar and outer-planar graphs are important as real-world networks (e.\,g., road-networks) are often close to planar and certain problems admit efficient solutions for planar graphs but not in general. The defining characteristic of these graph classes is that they admit a crossing-free drawing (for outer-planarity, every vertex is located on the outer face). There is a plethora of drawing algorithms for these graphs that can be used to create a visual certificate~\cite{DBLP:books/ph/BattistaETT99}. Scalability can be problematic 
and computing drawings really \emph{highlighting} planarity is an intriguing open question.


\subparagraph{Stack and Queue Number.} Stack and queue number are 
graph parameters arising for instance in the context of scheduling and VLSI layout (see~\cite{HR92,Yan89}). If a graph has a bounded stack (queue) number it admits a stack layout (queue layout, resp.) which uses few \emph{pages}, i.\,e., layers in which the edges are drawn. On each page 
in a stack layout, no edges cross while in a queue layout, no edges nest. These features can be easily detected, especially if one shows each page separately; i.\,e., a stack (queue) layout with $k$ pages visually proves that the stack (queue) number is \emph{at most} $k$. On the other hand, for this kind of assertion, user interaction may be required.

\subsection{Assertions that are Challenging to Visually Prove}
\label{sec:example:hard}

\subparagraph{$k$-Connectivity.}  Visually proving that a graph is $k$-connected is much more difficult compared to the complementary question discussed in~\cref{sec:examples:advanced}.
%
%
Specifically, the prosecution lawyer needs to highlight a sparsification $G'$ (i.\,e., a $k$-connected spanning subgraph) of $G$ which is $k$-connected, similar to the visual proof in~\cref{sec:connected}. 
We can utilize the Nagamochi-Ibaraki algorithm~\cite{NI}, which computes a $k$-connected spanning subgraph $G'$ with $O(kn)$ edges in linear time. 
Therefore, the judge may verify $k$-connectivity in $O(kn)$ time using $G'$, which is much faster for dense graphs with $O(n^2)$ edges.

However, for highly connected and large graphs (i.\,e., large values of $k$ and $n$), the perceptional complexity 
will be quickly increasing, making the design of effective visual proofs a significantly challenging problem.
For small values of $k$, say $k=2$, one may able to design more effective  visual proofs by highlighting the structure of the sparsification better. 
Note that the problems of determining planar~\cite{GOLDSCHMIDT199697} or minimum sized~\cite{DBLP:conf/ipco/CheriyanSS98} $2$-connected spanning subgraphs are NP-complete while there is a PTAS for the latter problem~\cite{DBLP:journals/tcs/GeorgiadisIK20}.


\subparagraph{Non-Planarity and Non-Outerplanarity.} It is well known that a graph is non-planar if and only if it contains either a $K_5$ or $K_{3,3}$ \emph{minor}~\cite{kuratowski} while it is non-outer-planar if and only if it contains either a $K_4$ or $K_{2,3}$ minor~\cite{AIHPB_1967__3_4_433_0}. Namely, a minor is a graph that can be obtained from the initial graph by a series of \emph{edge contractions}, i.\,e., one identifies both ends of an edge. As each vertex in the minor can correspond to large subgraphs of the initial drawing, it may be infeasible to visualize them in an easy-to-grasp fashion in a static layout. Thus, one may want to animate the edge contraction sequence. We leave it as an interesting open question whether such a visualization would be a convincing visual proof for non-planarity.

\subparagraph{Parameterized and CoNP-hard Assertions.} Other interesting variants of some problems presented in~\cref{sec:examples:easy,sec:examples:advanced,sec:examples:canonical} are \emph{parameterized assertions}. That is, instead of asking for a Hamiltonian cycle, we may ask for the existence of a cycle of length $k$. Similarly, 
we could be interested to prove that there  is 
an independent set, a clique or a dominating set of size $k$. Such assertions~can be easily visually proven, however, the perceptual complexity depends on the value $k$ as the judge has to count the size of the certificate.

The complementary questions for some of the presented problems are CoNP-complete; e.\,g.,  
showing the \emph{absence} of a Hamiltonian cycle, proving that the queue (stack) number of a graph is \emph{at least $k$}, and showing that a graph is \emph{not} $k$-colorable. 
This is in line with our conjecture that CoNP-complete problems admit no visual proofs.

\subsection{Visual Proofs for Network Analysis}
\label{sec:examples:network-analysis}

So far, we have mainly discussed how to visually prove assertions stemming from graph theory. In network analysis, other analysis are equally important. 
We now discuss how we can visually prove assertions in this domain.

\subparagraph{Diameter and Center.} 
The \emph{diameter} of a graph is the length of the longest shortest path between any two vertices in a graph, and the \emph{center} of a graph is a vertex whose distance (i.\,e., shortest path) to all other vertices in the graph is minimized. Both are fundamental properties of graphs based on the distance between the vertices. 
A visual proof of the diameter can be obtained by the small multiples of the level drawings of the BFS trees rooted at each vertex $v$, where the BFS tree with the maximum depth is highlighted as the diameter of the graph (in fact this single BFS tree is sufficient for a lower bound for the diameter). 
Such a level drawing of the BFS tree rooted at a vertex $v$ can be also used as a visual proof for vertices whose distance from $v$ is exactly $k$, by highlighting the vertices on level $k$. 
Similarly, it can be used for a visual proof for the graph center problem,
however this is more involved due to the comparison of the sum of all distances from $v$ to other vertices. 
The perceptual complexity of these visual proofs increases with size and density.



\subparagraph{Other Network Analysis Measures.} There is a plethora of important measures in network analysis, e.\,g. \emph{centrality}, \emph{$k$-core} and  \emph{automorphic equivalence}~\cite{WS}. The former two can be computed in polynomial time, while the latter is Isomorphism-complete. 
Possible visual proofs for these measures can be a level drawing or a radial drawing, where the level or concentric circle is defined based on the corresponding measures, such as centrality values or $k$-core index. 
Moreover, the $k$-core analysis can be visualized by a topographic map-style inclusion drawing, which may serve as a visual certificate.
Note that these visual proofs require a careful analysis by the judge resulting in linear perceptual complexity, that is, novel visual certificates that the judge can validate more efficiently are of interest.

\section{Limitations Related to Human Factors}
\label{app:human}

\subparagraph{Background Knowledge Required for Checkability.} In practical settings, checkability,  i.\,e., the judge's ability to  check the validity of an assertion efficiently based on their mental model, also depends on the background knowledge of the judge. For example, in \cref{sec:examples:easy}, we required knowledge of the terms cut-vertex, connectivity and Hamiltonian cycle. 
Thus, we may be tempted to require the judge to know basic notions of graph theory, specifically knowledge of  the properties for which the visual proof is exhibited. In addition to graph-theoretic background knowledge, familiarity with the chosen visualization style may be important. For instance, we have seen in \cref{sec:hamilton}
 that less frequently used paradigms such as adjacency matrix visualizations may allow for very efficient visual certificates. 
These aspects are limitations, when the audience is in fact non-expert and some required background knowledge is lacking. 
If so, additional explanations may also be presented to the audience examining the visual certificate, similar to court proceedings; see also \cref{sec:intro}.

\subparagraph{Subjective Aspects of Unimpeachability.}
In \cref{sec:certificate}, we introduced unimpeachability as a combination of information faithfulness and task readability. As a consequence, the defense lawyer has to check properties that can either be objectively fulfilled or violated. Thus, in its most simple form, the defense lawyer can be a computer program that checks whether the visualized graph is actually the input graph (such as in~\cref{ex:hamiltonian}) and the judge can rely on the fact that the defense lawyer would have rejected the drawing if it was not unimpeachable. Moreover, such a defense lawyer software may be implemented once for a certain drawing style and reused for other visual proof processes utilizing the same layout technique.
We believe that our objective definition of unimpeachability works  well in scenarios where we simply want to check that the drawing method worked correctly (as in~\cref{ex:hamiltonian}) or where the judge has no reason to distrust the prosecution lawyer (as in~\cref{ex:cut-vertex}). 
However, the requirements for unimpeachability are context dependent and a visual proof may need to be defended against  subjective counterarguments,  brought forth by an adversary (as in~\cref{ex:mst}) or by the skepticism of the judge (i.\,e., a single person fulfills both roles judge and defense lawyer). In such scenarios, there may be further subjective aspects of unimpeachability, e.\,g., the judge may expect the visualization to be similar to an already known layout of the graph, as otherwise the defense lawyer may question that the visual certificate shows the graph in question, i.\,e., there is feedback from the defense lawyer to the judge; see the dashed connection 
in~\cref{fig:teaser}. 

%

\subparagraph{Uncertainty related to the Mental Model.}

The judge's mental model is an abstraction that is difficult to describe as it can vary from judge to judge due to differences in both cognition \cite{Miyake2012} and perception~\cite{Grzeczkowski2017}. Further, it may even be uncertain to the judge how they abstract the visualization.
%
Thus, there 
are important open questions related to the judge's mental model: It is important to understand how visualization techniques can accurately influence the abstraction of the judge so that the evidence gathered by the prosecution lawyer can be translated into the mental model as unmodified as possible. Gathering empirical evidence that our concept of perceptual complexity captures the reality, i.\,e., that judges indeed use the mental model to establish the verdict, is an intriguing open problem. Moreover, we may ask which perceptual complexity is still accepted by human judges (linear workload may already be  overwhelming). Finally, it may be worth investigating alternative measures for perceptual complexity that are agnostic to whether or not a human judge actually follows a deterministic algorithm. Such measures may be similar to the ones used in predictive models in HCI such as KLM~\cite{DBLP:journals/cacm/CardMN80} or GOMS~\cite{DBLP:books/lib/CardMN83}.


\end{document}